\documentclass[prl, aps,
twocolumn,
longbibliography,
amsmath,amssymb,
floatfix
]{revtex4-2}

\usepackage{physics}
\usepackage{epsf}
\usepackage{float}
\usepackage{graphicx}
\usepackage{amsmath}
\usepackage[usenames]{color}
\usepackage{float}
\usepackage{bm}
\usepackage{relsize}
\usepackage[normalem]{ulem}

\def \be {\begin{equation}}
\def \ee {\end{equation}}
\def \bea {\begin{align}}
\def \eea {\end{align}}

\def \BEA {\begin{eqnarray}}
\def \EEA {\end{eqnarray}}
\def \BC {\begin{cases}}
\def \EC {\end{cases}}
\usepackage{siunitx}
\usepackage{tabularx}
\usepackage{xcolor}
\usepackage[version=4]{mhchem}
\usepackage{ctable}

\usepackage[unicode=true,colorlinks=true,citecolor=blue,urlcolor=blue]{hyperref}

\def\be {\begin{equation}}
\def\ee {\end{equation}}
\def\bea {\begin{align}}
\def\eea {\end{align}}

\def\bee{\begin{eqnarray}}
\def\eee{\end{eqnarray}}
\def\BC {\begin{cases}}
\def\EC {\end{cases}}

\makeatother

\begin{document}
\title{Circular polarization immunity of the cyclotron resonance photoconductivity in two-dimensional electron systems}

\author{E.~Mönch$^1$, P.~Euringer$^1$, G.-M. Hüttner$^1$, I.~A.~Dmitriev$^1$, D.~Schuh$^1$, M.~Marocko$^1$, J.~Eroms$^1$, D.~Bougeard$^1$, D.~Weiss$^1$ and S.~D.~Ganichev$^1$}

\affiliation{$^1$Terahertz Center, University of Regensburg, 93040 Regensburg, Germany}

%\affiliation{$^2$ Rzhanov Institute of Semiconductor Physics, SB RAS, Novosibirsk, 630090 Russia}

%\affiliation{$^2$Ioffe Institute, 194021 St. Petersburg, Russia}

%\affiliation{$^3$CENTERA, Institute of High Pressure Physics PAS, 01142 Warsaw, Poland}

%\affiliation{$^3$ International Center for ss Nanoarchitectonics, National Institute of Material Science, 1-1 Namiki, Tsukuba 305-0044, Japan}

%\affiliation{$^4$ Research Center for Functional Materials, National Institute of Material Science, 1-1 Namiki, Tsukuba 305-0044, Japan}

\date{\today}

\begin{abstract}	
			Studying the cyclotron resonance (CR)-induced photoconductivity in GaAs and HgTe two-dimensional electron structures, we observed an anomalous
			photoresponse for the CR-inactive geometry being of almost the same magnitude as the CR-active one. This observation conflicts with simultaneous transmission measurements and contradicts the conventional theory of CR which predicts no resonant response for the CR-inactive geometry. We provide a possible route to explain this fundamental failure of the conventional description of light-matter interaction and discuss a modified electron dynamics near strong impurities that may provide a local near-field coupling of the two helicity modes of the terahertz field at low temperatures. This should result in a CR-enhanced local absorption and, thus, CR photoconductivity for both magnetic field polarities. 
\end{abstract}

\maketitle

Cyclotron resonance is a fundamental and well established textbook phenomenon widely used in solid state research to study energy dispersion, scattering times, energy structure of excitons and impurities as well as other electronic properties of bulk and two-dimensional systems (2DES)~\cite{Seeger2004}. Under the condition of CR in a magnetic field $B$, an electromagnetic wave with frequency $f = q B/2\pi m$ is resonantly absorbed by the conduction carriers with charge $q$ and mass $m$, accelerated on spiral or circular cyclotron orbits. Basic approaches to investigate CR include (i) radiation transmission/reflection, (ii) photoconductivity, and (iii) quenching of the photoluminescence due to CR absorption, for review see~\cite{Kono2012}.
%~\cite{Baranov1977, Romestain1980}
%Baranov1977 = P.G. Baranov, Yu.P. Veshchunov, R.A. Zhitnikov, N. G. Romanov, and Yu.G. Shreter, Pis'ma Zh. Eksp. Teor. Fiz. 26, 3699 (1977) [JETP Lett. 26, 249 (1977)].;\\ Romestein1980= R. Romestain and C. Weisbuch, Phys. Rev. Lett. 5, 2067 (1980);\\ 
%Kono2001=J. Kono, \textit{Cyclotron Resonance}, in: \textit{Methods in Materials Research}, edited by E. N. Kaufmann, R. Abbaschian, A. Bocarsly, C.-L. Chien, D. Dollimore, B. Doyle, A. Goldman, R. Gronsky, S. Pearton, and J. Sanchez (John Wiley \& Sons, New York, 2001),  Chap. 9b.2.

The transport approach using microwave and terahertz (THz) photoconductivity/photoresistance, originally termed the cross-modulation method, has been suggested for CR studies by Zeiger et al. in 1958~\cite{Zeiger1958}. An advantage of this technique is that the sample itself acts as a detector. Thus, it can be applied even to micrometer-size structures where reliable CR transmission and reflection measurements are impossible. Since the first observation of the CR in 1953 it is known that, for circularly polarized light propagating along or against the magnetic field, CR is only possible if the helicity of light and the sense of cyclotron motion match. Indeed, resonant acceleration requires that the light’s electric field rotates synchronously and in the same (opposite) sense as the positively (negatively) charged carriers undergoing a cyclotron motion. The strong $B$-asymmetry of CR controlled by the wave helicity (CR-active/inactive polarity of $B$) is confirmed by many CR experiments and is widely used to determine the type of conduction carriers.

In sharp contrast to this textbook behavior, our present experiments using several GaAs- and HgTe-based 2DES reveal that the CR photoconductivity becomes insensitive to the radiation helicity when the measurement temperature $T$ is lowered to that of liquid helium or below: The amplitude of the CR signals excited by a circularly polarized THz radiation is observed to be almost the same for both CR-active and inactive polarities of $B$, and for both helicities. Strikingly, the conventional behavior of the photoconductivity, i.e., the CR present for the active polarity only, is gradually restored at higher $T$. Unlike the anomalous photoconductivity, simultaneous measurements of CR in transmission show the ordinary helicity dependence at all $T$. 

Our experiments, performed on large structures with lateral size strongly exceeding the THz laser spot, unambiguously demonstrate that: (i) the helicity-insensitive photoresponse can be detected in CR photoconductivity which directly reflects resonant CR absorption and associated heating of electrons; (ii) the anomaly disappears at higher temperature, (iii) the anomaly is not present in transmission, and (iv) the observed immunity is not related to external factors, like antenna effects or diffraction at the metallic parts of the experimental setup, contacts, or sample's edges. We argue that the observed CR anomalies in the photoconductivity can be attributed to an enhanced near-field absorption and suppressed reflection of THz radiation in the vicinity of rare strong impurities or inhomogeneities in 2DES.
%rare strong impurities/defects which should be treated beyond the conventional Born approximation and can be present in the investigated GaAs and HgTe structures. 
Such near-field effects can locally destroy the rotational and translational symmetries of electron transport leading to strong mixing of both helicity modes in the dynamically screened non-uniform THz field acting on 2D electrons. These effects are not detectable in transmission measured in the far field at a large distance from the sample, and can be suppressed at high $T$ where electron transport is dominated by electron-phonon scattering. Within this interpretation, the CR photoconductivity in response to a circularly polarized THz radiation can serve an indispensable tool to test the technological quality and nature of disorder in 2DES. %high-mobility electron systems.

%\begin{table*}
%	\centering
%	\begin{tabularx}{\textwidth}{XXXXXXX}
%		\toprule[0.05cm]\addlinespace[0.2cm]
%		Sample& QW thickness $d$ (nm) & geometry & substrate orientation & $n_s$ (10$^{11}$~cm$^{-2}$) & $\mu$ ($10^4$~cm$^2$/Vs) \\\midrule[0.025cm]\addlinespace[0.1cm]
%		\#A&  8.1 	&  van der Pauw	& (013) & 3.5 & 1.9 \\
%		\#B&  5.7 	&  van der Pauw	& (\re{???}) & 1.3 & 0.05 \\
%		\addlinespace[0.1cm]\bottomrule[0.05cm]
%	\end{tabularx}	
%	\caption{Transport characteristics of sample\#A and \#B. For the latter the parameters were obtained after 10 min of LED illumination.}
%	\label{TabS1}
%\end{table*}

\begin{figure}
	\centering \includegraphics[width=\linewidth]{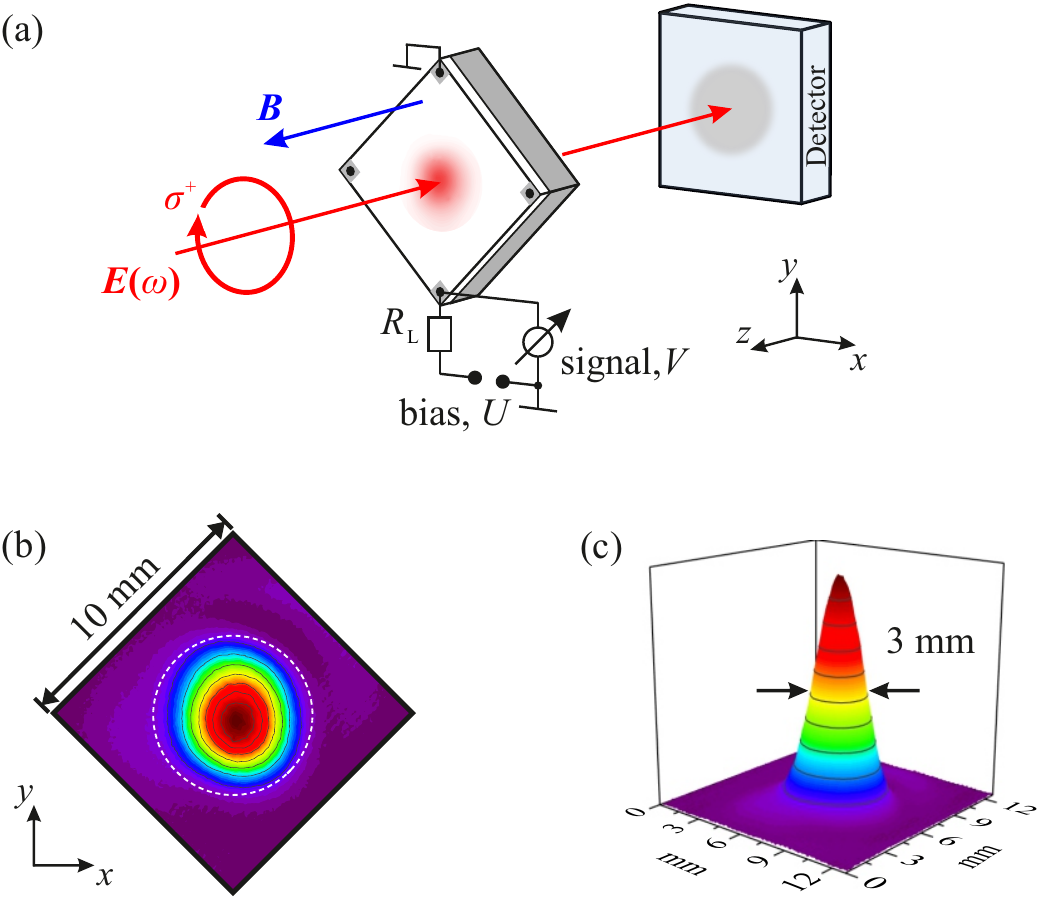}
	\caption{(a) Sketch of the experimental setup. 
	%{The transmitted power of a normally-incident circularly-polarized THz laser beam was recorded with a low-noise pyroelectric detector as a function of magnetic field $B$ applied normal to the 2DES plane. Simultaneously, the radiation-induced change of resistance (i.e., the photoresistance $\Delta R$) was measured using ohmic contacts at the corners, by applying a dc bias of opposite polarities or a low-frequency ac bias over a load resistor, $R_{\rm L}$.}
		(b) and (c): Measured intensity profile of the $f = 0.69$~THz beam focused at the 10$\times$10~mm$^2$ GaAs sample. The edges of the sample are not irradiated. Dashed circle in (b): Spot diameter of 6~mm at $13.5\%$ of the peak intensity. Arrows in (c): Full width at half maximum. 
		%The white dashed circle in (b) marks the spot diameter ($\approx$ 6~mm) at the level of $\mathrm{e}^{-2}\approx13.5\%$ $\mathrm{e}^{-2}$ of the peak intensity. The full width at half maximum of $\approx$ 3~mm in (c) is marked by arrows. 
	}
	\label{FigS1} 
\end{figure}

%Sample numbers: 
%
%- HgTe
%	Sample #A = 170206
%	Sample #B = 180406
%
%	10.10.2021
%
%- GaAs
%	Sample #A = C150506B (MIRO Paper: Sample #A); 
%	Sample #B = C150512C (MIRO Paper: Sample #D)
%	Mobility about 10^6

%Sample numbers: 
%	
%	25.01.2022 
%	
%	Sample #1_D = C150512C (GaAs)
%	Sample #2_A = C150506B (GaAs)
%	
%	Sample #3 	= 170206	(HgTe 8.1nm)
%	Sample #4	= 180406	(HgTe 5.7nm)
%
%	Sample #5	= Stack #2 Marina (MLG)
%

 We studied several 2DES including AlGaAs/GaAs quantum wells (QWs) (van der Pauw $10\times10$~mm$^2$ samples), HgCdTe/HgTe QWs of 8.1 and 5.7~nm thickness ($7\times7$~mm$^2$), and a high quality hBN-encapsulated monolayer graphene (MLG) (Hall bar, 24$\times$5~$\mu$m$^2$). Magnetotransport measurements at $T = 4.2$~K yielded electron densities from 1 to 20$\times$10$^{11}$~cm$^{-2}$ and  mobilities from 2 to 10$\times 10^5$~cm$^2$/V\,s (GaAs) and 1 to 2$\times 10^4$~cm$^2$/V\,s (HgTe). For technological design and magnetoresistance data, see Ref.~\cite{SM}.

\begin{figure*}
	\centering \includegraphics[width=\linewidth]{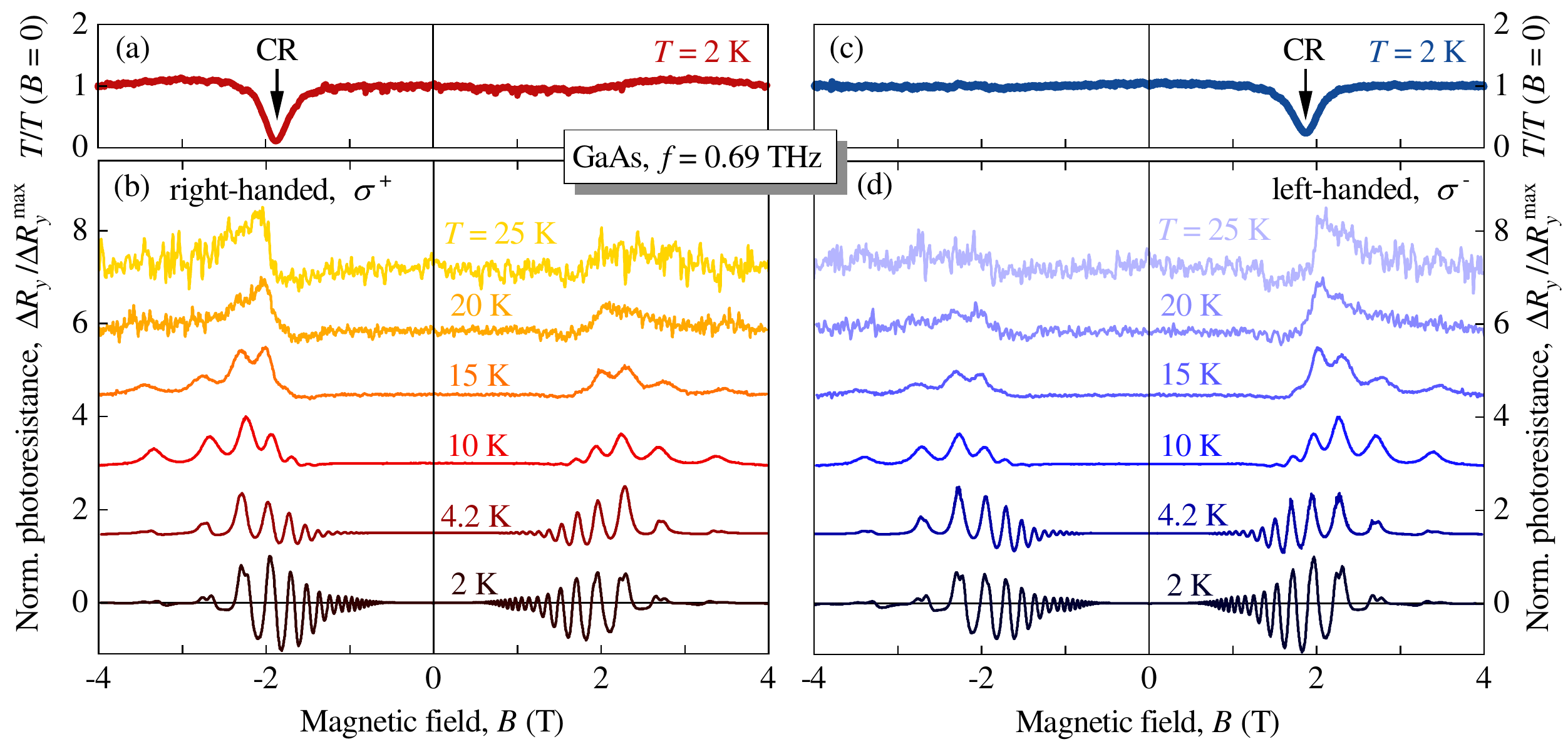}
	\caption{Normalized transmittance, $T(B)/T(0)$, measured at $T=2$~K on the GaAs sample \#1 for right-handed (a) and left-handed (c) $f = 0.69$~THz radiation. Black arrows mark the positions of CR. (b) and (d):  the corresponding photoresistance, $\Delta R_y$, normalized to its maximum value, $\Delta R^{\rm max}_y$, measured at $T$ from 2 to 25~K. Traces are shifted by 1.5 for clarity.}
	\label{FigR1} 
\end{figure*}

To study the CR, the samples were placed in an optical cryostat with $z$-cut crystal quartz windows covered by black polyethylene films to avoid uncontrolled illumination by ambient light. An optically pumped continuous wave molecular gas laser~\cite{Dantscher2017} operating at $f = 0.69$, 1.63 and 2.54~THz provided normally incident radiation parallel to applied magnetic field, see Fig.~\ref{FigS1}(a). 
%(photon energies $\hbar\omega$ = 2.9, 6.7 and 10.5~meV, respectively)}. 
The intensity distribution of the THz laser beam focused on the sample center was measured by a pyroelectric camera, Figs.~\ref{FigS1}(b) and (c), revealing a nearly Gaussian profile with spot diameters $d = 6$, 4.2, and 2.8~mm at the level of $\mathrm{e}^{-2}\approx13.5\%$ of the maximum intensity (few W/cm$^2$). This assured a negligible contribution of the sample edges and contacts in both photoresistivity and transmission obtained on QW samples. Right- ($\sigma^+$) and left- ($\sigma^-$) handed circularly polarized radiation was produced by $x$-cut crystal quartz quarter wave plates.  % placed behind the sample, see Fig.~\ref{FigS1}(a). 
To measure the photoresistance $\Delta R$ (i.e., the radiation-induced change of resistance) in QW structures,
either a dc or ac bias voltage $U$ was applied over a load resistor to point contacts at opposite corners of the sample; simultaneously, radiation transmission was measured with a pyroelectric detector, see Fig.~\ref{FigS1}(a). 
In graphene, the bias voltage was applied to two end terminals of the Hall bar structure. The signal $V$ was measured using standard lock-in technique. The photoresistance was extracted either by subtracting signals %obtained 
for opposite polarities of the dc bias, or by applying the double modulation technique \cite{Kozlov2011, Otteneder2018,SM}.

\begin{figure}[t!]
	\centering \includegraphics[width=\linewidth]{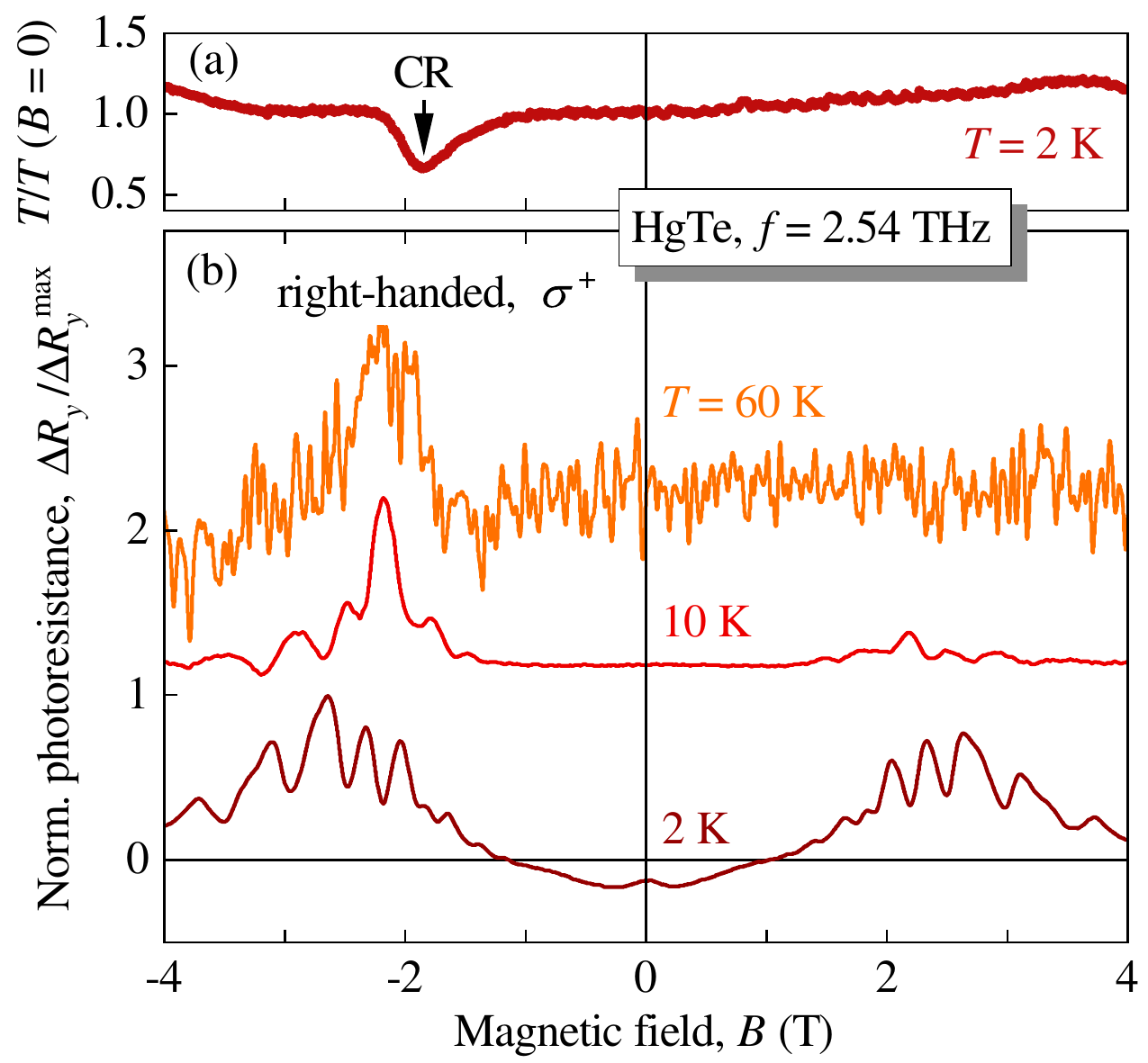}
	\caption{Normalized transmittance at $T=2$~K (a) and normalized photoresistance at $T=2$, 10, and 60~K (b) measured on the HgTe sample \#3 with  8.1~nm QW width for right-handed $f = 2.54$~THz radiation. Black arrow marks the position of CR. $\Delta R$ traces are shifted by 2 for clarity.}
	\label{FigR3} 
\end{figure}

\begin{figure}[t!]
	\centering \includegraphics[width=\linewidth]{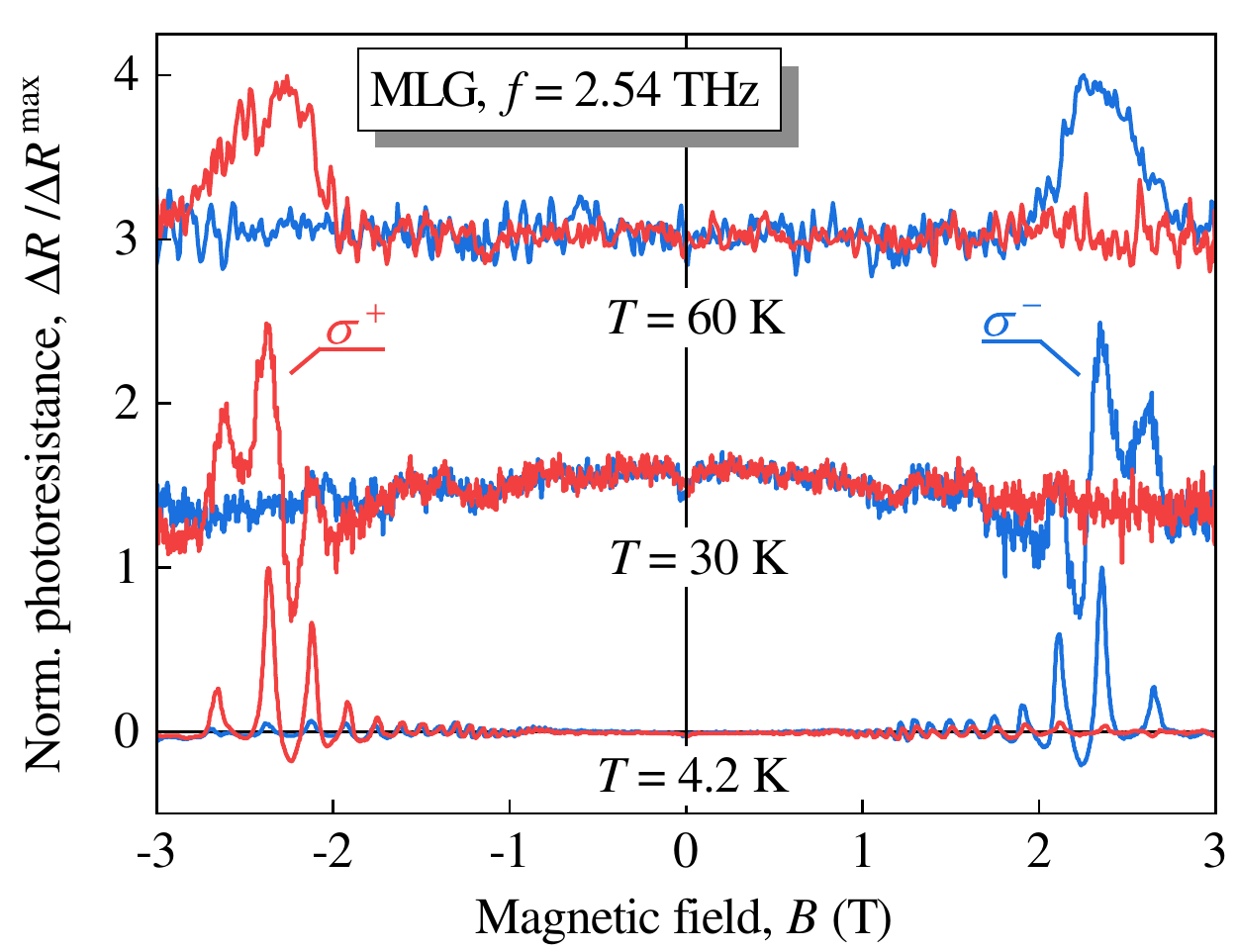}
	\caption{Normalized photoresistance measured at $T=4.2$, 30, and 60~K on the monolayer graphene sample \#5 with $2.1\times 10^{12}$~cm$^{-2}$ carrier density for right-handed (red) and left-handed (blue) $f = 2.54$~THz radiation. Traces are shifted by 1.5 for clarity. Transmittance measurements were not feasible due to small size of the MLG sample.
	}
	\label{FigR4} 
\end{figure}

The CR was clearly detected in both transmission and photoresistance in a wide temperature range between 2 and 90~K. Typical results for GaAs, HgTe and MLG samples are shown in Figs.~\ref{FigR1}, \ref{FigR3}, and \ref{FigR4}, respectively, and confirmed by measurements at other frequencies and samples~\cite{SM}. As expected, the transmission traces (top panels in Figs.~\ref{FigR1}, \ref{FigR3}) clearly show that the CR excited by circularly polarized radiation appears for one $B$-polarity only ( CR-active polarity, $B<0$ for $\sigma^+$ helicity and $B>0$ for $\sigma^-$ helicity). Strong wide dips at the position of CRs in transmission (arrows) are produced by resonant reflection and absorption of radiation, the latter mechanism playing a minor role in high-mobility and high-density 2DES studied here \cite{Abstreiter1976, Chiu1976, Mikhailov2004, Zhang2014, Herrmann2016, Dmitriev2012,Savchenko2021,Savchenko2022}. 
Importantly, the transmission remained the same at all $T$ showing no resonant features on the CR-inactive side under all conditions.

At the CR-active side, photoresistance traces (bottom panels in Figs.~\ref{FigR1}, \ref{FigR3} and Fig.~\ref{FigR4}) display the well established behavior associated with resonant electron gas heating under CR absorption. At high $T$, we observe a single CR peak in $\Delta R$ caused by heating-induced decrease of the electron mobility \cite{Ganichev2005}. 
%Further references: Muraviev2013, Freitag2012, Betz2012, Heyman2015, Ryzhii2019, Jago2019 
At low $T$, the heating under CR reduces the amplitude of the Shubnikov-de Haas oscillations (SdHO) reflecting their exponential sensitivity to electron temperature. SdHO are completely suppressed at high $T$, while at low $T$ the photoresistance shows $1/B$-oscillations with the period of SdHO. These oscillations are resonantly enhanced near CR where the heating is maximized. Traces at intermediate temperatures demonstrate a combined effect of CR heating on SdHO and mobility.  

While the results on the CR-active side are common, the CR-inactive side clearly shows an anomalous behavior. Strikingly, at low $T$ the CR-enhanced photoresistance in Figs.~\ref{FigR1}, \ref{FigR3} has almost the same magnitude for positive and negative $B$, as well as for both helicities  -- a result which one would expect not for circular but rather for linear polarization where both circular components have equal weights. Furthermore, the relative magnitude of the signal on the CR-inactive side becomes progressively weaker at higher measurement temperatures, and disappears at the highest $T$ thus restoring the behavior expected for circular polarization. Importantly, the $T$-dependence of the anomaly excludes any extrinsic mechanisms related to possible breakdown of the circular polarization in the optical setup,
%which may spoil the circular polarization in the optical setup,
%related to possible deterioration of 
and shows that it has an intrinsic origin reflecting an anomalous high-frequency current response to the THz driving inside the 2DES. The large size of GaAs and HgTe samples in comparison to the THz beam spot  
excludes the influence of samples' edges and contacts. Moreover, the CR photoresistance measured on a small MLG sample, where illumination of the edges and contacts could not be avoided, shows regular CR behavior with no resonant features on the CR-inactive side, see Fig.~\ref{FigR4}. This demonstrates that the helicity anomaly in the CR absorption is not universal but rather reflects peculiarities of the dynamic response in specific 2DES.

The observed polarization immunity suggests that the conventional description of light-matter interaction in terms of local dynamic conductivity $\hat{\sigma}(\omega, q\to 0)$ is not applicable. Within this standard approach, the electric field $\mathbf{E}$ of a plane circularly polarized wave, acting on electrons in an isotropic uniform 2DES, induces a uniform circular electric current $\mathbf{j}=\hat{\sigma}\mathbf{E}$ of the same helicity. Both quasiclassical and quantum kinetic theory predict that this current should be resonantly enhanced at the CR at positive or negative $B$ only, depending on helicity~\cite{Seeger2004, Kono2012, Dmitriev2012}. As long as both 2DES and THz field remain uniform and isotropic, there is no coupling between the two helicity modes and thus such description is incompatible with the polarization immunity.

To overcome this apparent paradox, we thus need to consider some intrinsic source of broken translational or rotational symmetry leading to mixing of the otherwise independent helicity modes. A plausible resolution comes from understanding that the standard approach assumes uncorrelated scattering by a large number of weak impurities which, after disorder averaging, yields full description in terms of uniform and isotropic $\sigma(\omega, q\to 0)$. However, the electron flow can be essentially modified near rare strong impurities or inhomogeneities %of microscopic size
\cite{Baskin1978,Bobylev1995,Mirlin2001,Dmitriev2012,Dmitriev2004,Beltukov2016,Dorozhkin2017,Chepelianskii2018}. Near such impurities the system is neither translationally invariant nor isotropic leading to strong coupling between the two helicity modes and thus enabling the polarization immunity. Taking into account that in a uniform high-density and high-mobility 2DES the most part of radiation is reflected in the vicinity of CR \cite{Abstreiter1976, Chiu1976, Mikhailov2004, Zhang2014, Herrmann2016}, a 2DES with such strong impurities can be visualized as an old mirror with dark spots: Near impurities the 2DES is ``dirty'' and does not reflect the THz wave effectively. Therefore, the near field acting on the electrons is stronger. This yields double enhancement of absorption due to stronger scattering and stronger field at the ``dark spots'' which, therefore, are also ``hot spots''. Under such conditions, one faces the nontrivial task to self-consistently calculate inhomogeneous local currents induced by the external uniform THz field and corresponding dynamic screening of the external field by the 2DES before averaging over disorder. Moreover, the strongly nonuniform screened THz field implies local excitation of short-wavelength plasmons and possible viscous effects. Within this interpretation, the conventional behavior of CR can be restored at high $T$ due to the increasing role of phonon and electron-electron scattering. Furthermore, the evanescent waves associated with the THz near field of opposite helicity emerging near strong impurities should not affect conventional CR in transmission, detected in the far field at a large distance from the sample. The non-universality of the proposed mechanism, relying on the presence of strong impurities, is supported by the CR photoresistance in MLG showing no helicity anomalies at all $T$.
%is supported by the fact that in MLG the CR photoresistance shows no helicity anomalies 
%and remains ordinarily helicity-sensitive 
%in the whole investigated temperature range.   

The idea that strong impurities can strongly modify electron transport in 2DES is not new. It was extensively studied in both static and dynamic regimes in the context of non-Markovian classical memory effects, see, e.g. \cite{Baskin1978,Bobylev1995,Mirlin2001,Dmitriev2012,Dmitriev2004,Beltukov2016,Dorozhkin2017,Chepelianskii2018}. 
These studies, however, have mostly concentrated on electron transport in the presence of uniform dc or ac electric fields. Apart from Ref.~\cite{Chepelianskii2018}, discussing similar ideas, they did not considered the possibility of strong back action of inhomogeneous currents on the field acting on electrons discussed above.

\begin{figure}
	\centering \includegraphics[width=\linewidth]{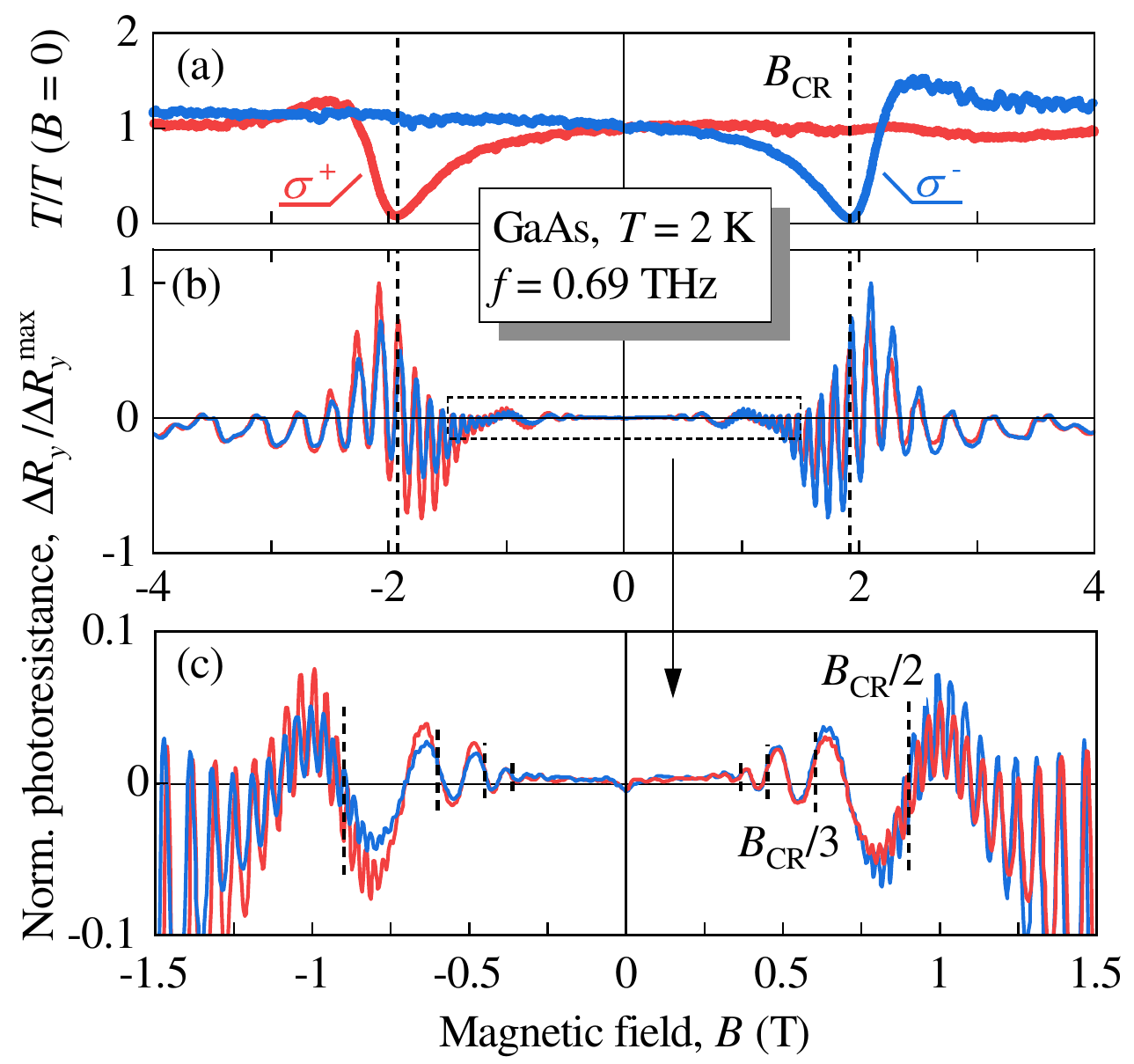}
	\caption{Normalized transmittance (a) and photoresistance (b) measured at $T=2$~K on the GaAs sample \#1 for right-handed (red) and left-handed (blue) $f = 0.69$~THz radiation after brief illumination by room light prior to measurements. The dashed vertical lines labelled with $B_{\rm CR}$ mark the position of the CR. (c): Enlarged low-$B$ data from marked area in (b) %. Room light illumination leads to emergence of 
		showing MIRO with nodes at the CR harmonics (dashed lines).}
	\label{FigR2} 
\end{figure}

Before concluding, we shortly address polarization anomalies previously detected \cite{Smet2005,Herrmann2016} in studies of microwave-induced resistance oscillations (MIRO) \cite{Zudov2001,Mani2002,Zudov2003,Dmitriev2012}, magnetooscillations in photoresistance coupled to the  harmonics of the CR. In Fig.~\ref{FigR2} we present transmission and photoresistance data for the same GaAs sample as in Fig.~\ref{FigR1} but now obtained after brief illumination by room light 
%at $T=2$~K 
prior to measurements. 
%In this type of 2DES illumination typically 
This results in an increasing electron density and mobility due to the persistent photoconductivity effect~\cite{Mooney1990,SM}. %T. N. Theis Comm. Cond. Mat. Phys. 16, 167, 1992.
Here, the transmission still shows regular helicity dependence with no features on the CR-inactive side. Similar to Fig.~\ref{FigR1}, the photoresistance in Fig.~\ref{FigR2}(b) shows an almost complete immunity to the helicity of the THz wave. %{But now, after illumination,} 
However, in addition to the SdHO-related oscillations, 
%a new set of 
new magnetooscillations %\sdg{with lower frequency} 
appear with nodes at the CR, $B_{\rm CR}$, and its harmonics $B_{\rm CR}/2$, $B_{\rm CR}/3$, $\ldots$, see dashed lines in Fig.~\ref{FigR2}(c), which correspond to the %well-known 
MIRO effect~\cite{Zudov2001,Mani2002,Zudov2003,Smet2005,Dmitriev2012,Herrmann2016}.

While the detected helicity immunity of MIRO reproduces previous results \cite{Smet2005,Herrmann2016}, our findings provide a completely different perspective on its origin. Our findings contradict the commonly accepted view that the immunity is specific to MIRO or is related to technical details of measurements like illumination of contacts and sample edges. We demonstrate that the immunity has fundamental character and its origin is not linked to MIRO which just provide an alternative way to detect the helicity anomaly in THz or microwave absorption \footnote{Note that observation of MIRO also rules out the photo-ionization of impurities~\cite{Putley1966} as a mechanism for the resonant photoconductivity.}. 
Furthermore, the latest experiments of the Vienna group \cite{Savchenko2022} show that the helicity immunity of MIRO is also not universal.
%and, apparently, depends on details of disorder in a particular 2DES. 
Using a GaAs 2DES with similar density and mobility but with different technological design, they %rather 
observed the expected \cite{Dmitriev2012} strong helicity dependence of MIRO under sub-THz illumination %accurately 
reproducing the regular CR lineshape of the Drude absorption ~\cite{Savchenko2022}. 
%The regular helicity dependence of MIRO was also observed in illuminated 2DES on the surface of liquid He \cite{Zadorozhko2018}.

To summarize, our observations point to a fundamental failure in the conventional description of light-matter interaction in 2DES that shows up in a puzzling polarization immunity in CR-enhanced THz absorption. This behavior is in sharp contrast to the usual strong helicity dependence detected in simultaneous transmission experiments. We find that the observed anomaly is not universal and does not show up in all systems. Moreover, the helicity anomaly disappears at elevated temperatures restoring the conventional textbook CR behavior. We propose that the anomaly is related to a modified electron dynamics near strong impurities/inhomogeneities which can locally couple two otherwise independent helicity modes leading to near field CR-enhanced absorption for both magnetic field polarities. 

The authors thank A. Pimenov, M. L. Savchenko, A. Shuvaev, and J. H. Smet  for valuable discussions. We thank S.A. Dvoretsky and N.N. Mikhailov for fruitful discussions and growth of high quality HgTe QW wafers. We are grateful to K. Watanabe and T. Taniguchi for providing high-purity hBN crystals~\cite{Taniguchi2007}. We acknowledge the financial support of the Deutsche Forschungsgemeinschaft (DFG, German Research Foundation) via Project-ID 314695032 – SFB 1277 (Subprojects A01, A04, and A09) and via grant DM~1/5-1 (I.A.D.), of the Volkswagen Stiftung Program (97738), and of the IRAP  Programme  of  the Foundation   for   Polish Science   (S.D.G., grant   MAB/2018/9, project CENTERA).

\bibliography{all_lib}

\section{Supplemental Material}

\subsection{Technological design of the samples}
\label{Appendix1}

%Sample numbers: 
%	
%	25.01.2022 
%	
%	Sample #1_D = C150512C (GaAs)
%	Sample #2_A = C150506B (GaAs)
%	
%	Sample #3 	= 170206	(HgTe 8.1nm)
%	Sample #4	= 180406	(HgTe 5.7nm)
%
%	Sample #5	= Stack #2 Marina (MLG)

\subsubsection{AlGaAs/GaAs heterostructures}

%\begin{table*}
%	\centering
%	\begin{tabular}{|c|c|c|c|c|c|c|c|c|c|c|}
%		\hline
%		Sample       & QW thickness   &$n_e$(old) &$\mu$(old) & $\tau_p$(old) &   $n_e$(new) &$\mu$(new) & $\tau_p$(new) & $n_e$(new, rl) &$\mu$(new, rl)  & $\tau_p$(new, rl)   \\
%		& (nm) &10$^{11}$\,(cm$^{-2}$) & 10$^3$\,(cm$^2/$Vs) & (ps) & 10$^{11}$\,(cm$^{-2}$) & 10$^3$\,(cm$^2/$Vs) & (ps) & 10$^{11}$\,(cm$^{-2}$) & 10$^3$\,(cm$^2/$Vs) & (ps)\\
%		\hline
%		\#1$_{\rm D}$
%		C512C (GaAs)
%		& 10 & 24  &   980   & 33   &7 & 370 & 13 & 12 & 660 & 24  \\
%		\#2$_{\rm A}$
%		C506B (GaAs)
%		& 10 & 12  &  820    & 12.0   & 7 & 200 & 7 & 12 & 430 & 15   \\
%		\#3
%		170206 (HgTe)
%		& 8.1 & -  &  -    & -   & 3.6 & 18 & 0.3 & - & - & -   \\
%		\#4
%		180406 (HgTe)
%		& 5.7 & -  &  -    & -   & 1.3 & 9 & 0.13 & - & - & -   \\
%		\#5
%		Stack#2 Marina (MLG)
%		& - & -  &  -    & -   & 21 & 240 & 4.5 & - & - & -   \\
%		\hline
%	\end{tabular}
%	\caption{Sample parameters and transport data obtained at $T=2$~K, including the electron density $n_e$,  mobility $\mu$ and momentum relaxation time $\tau_p$. Parameters with the addition (old) were measured in corbino geometry taken from \cite{Herrmann2016}. The parameters with the addition (new) were obtained for the same samples, but measured in a van der Pauw configuration. Data with the abbreviation rl were measured with room light illumination. For Sample\#5 (MLG) the parameters are given for the highest back gate voltage.}
%	\label{sample}
%\end{table*}

The AlGaAs/GaAs quantum well (QW) structures  were grown by molecular-beam epitaxy (MBE) on 350~$\mu$m thick (001)-oriented GaAs substrates. The cross-section of the structure is shown in Fig.~\ref{FigA1}. First a buffer of a 100~nm thick GaAs layer was grown on top of the GaAs substrate. As next, to bind impurities and defects in the substrate while keeping them away from the active layers, an AlGaAs/GaAs-superlattice was grown (100 periods of 7~nm AlGaAs and 3~nm GaAs). Subsequently, a 10~nm wide GaAs QW was deposited on the superlattice, which was coated on the bottom side with 85~nm / 100~nm AlGaAs, and on the top side embedded in 30~nm / 45~nm AlGaAs. Here and below 
the thicknesses and distances were presented in this form $d_1 / d_2$ corresponding to sample \#1$_{\rm D}$ / sample \#2$_{\rm A}$. The AlGaAs layers had 32\% of Al content. Both samples were covered with 10~nm GaAs to protect them from oxidation processes. On both sides of the QW a Si-delta dopant was symmetrically introduced at 10~nm / 25~nm. The dopant concentration was also chosen in such a way that the QW remains symmetrical~\cite{Lechner2009}. In order to prevent segregation of Si, the growth temperature was lowered to \SI{500}{\degreeCelsius} in the doping region, while the rest of the sample (in particular the QW) was grown at \SI{570}{\degreeCelsius}. The growth temperature of the buffer layers was \SI{620}{\degreeCelsius}.

From the grown structures we prepared two 10$\times$10 mm$^2$ square-shaped samples \#1$_{\rm D}$ and \#2$_{\rm A}$ and equipped them with ohmic contacts at the corners. Samples were fixed in sample sockets and the contacts were connected to the socket pins via gold wires. The subscripts in the sample notation~\#1$_{\rm D}$ and \#2$_{\rm A}$ point to the fact that samples D and A made from the same corresponding wafers were studied in our previous paper \cite{Herrmann2016}.

%Angaben der Form xxxx / yyyy bedeuten: für Probe C150506B / C150512C
%Generell: AlGaAs heisst 32% Al Anteil in der Schicht.
%
%Beide Proben sind auf einem 350µ dicken GaAs Substrat gewachsen.
%Zunächst kommt ein Puffer aus einer 100nm dicken GaAs Schicht 
%und einem AlGaAs/GaAs 
%-Übergitter um Verunreinigungen und Fehlstellen im Substrat
%zu binden und von den aktiven Schichten fernzuhalten. 
%
%Anschließend wurde ein 10nm breiter
%GaAs QW gewachsen, der substratseitig in 100nm / 85nm AlGaAs, und zur Oberfläche hin
%in 45nm / 30nm AlGaAs eingebettet ist. Beide Proben sind mit 10nm GaAs abgedeckt um
%sie vor Oxidation zu schützen.
%
%Zu beiden Seiten des QWs ist symetrisch je eine Si-Delta-Dotierung im Abstand von  25nm / 10nm
%eingebracht. Die Dotierkonzentration wurde dabei ebenfalls so gewählt, dass 
%der QW symmetrisch bleibt. 
%
%Um eine Segregation von Si zu verhindern wurde die Wachstumstemperatur
%im Bereich der Dotierung auf 500°C abgesenkt, der Rest der Probe (insbesondere der QW)
%wurde mit 570°C gewachsen. Die Wachstumstemperatur der Pufferschichten betrug 620°C.

\subsubsection{HgCdTe/HgTe heterostructures}

\begin{figure}[h!]
 	\centering \includegraphics[width=\linewidth]{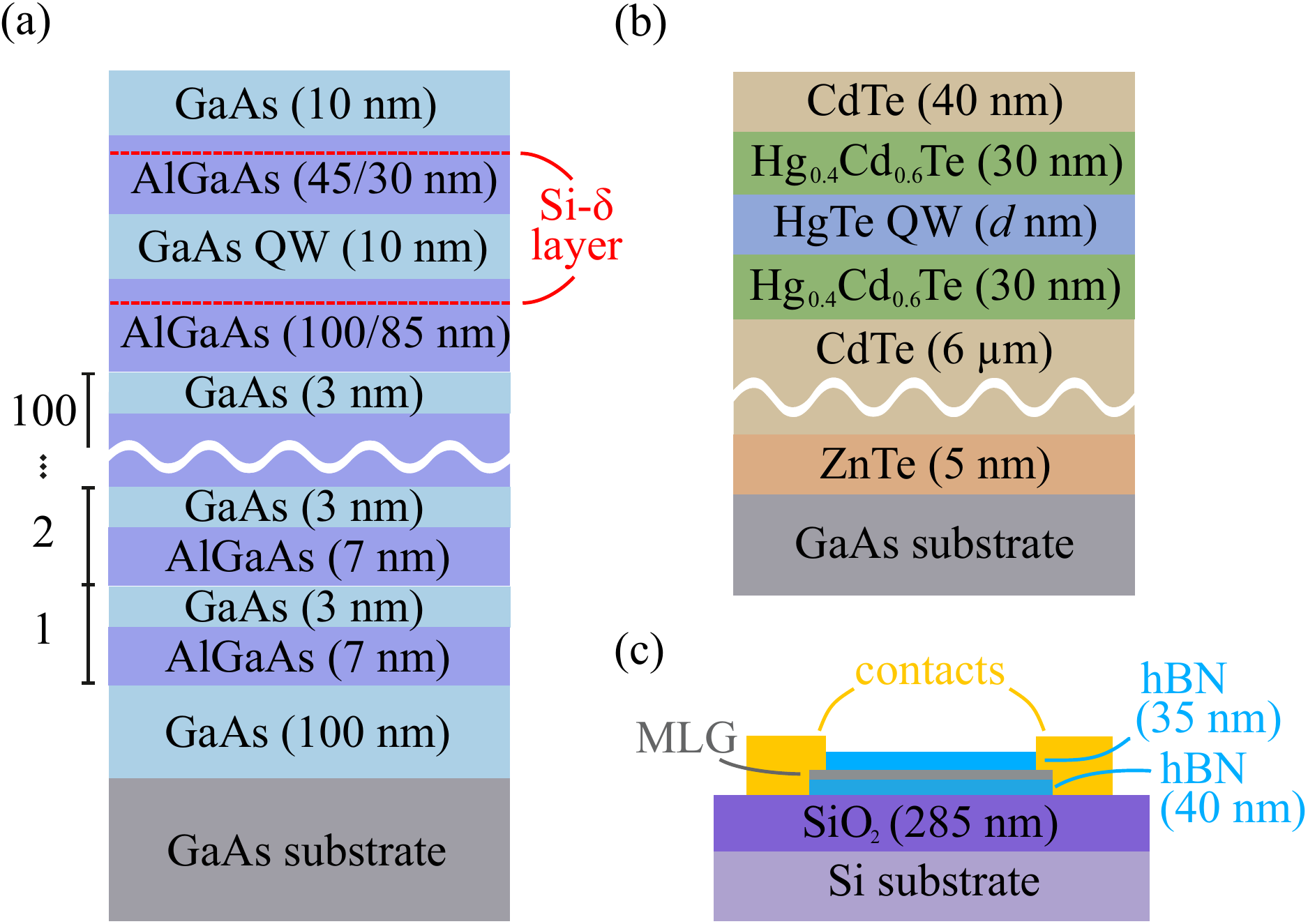}
 	\caption{Cross-sections of the (a) AlGaAs/GaAs heterostructures, (b) HgCdTe/HgTe heterostructures, and (c) hBN-encapsulated monolayer graphene. 
 		%The AlGaAs/GaAs QWs feature a thickness of 10~nm and a two Si-$\delta$-doping layers fabricated symmetrically around the 2DES (10~nm distance to the QW for sample\#1$_{\rm D}$ and 25~nm for sample\#2$_{\rm A}$), to increase its quality. The CdHgTe/HgTe heterostructures were grown on a GaAs substrate with QW above (sample\#3 with 8.1~nm) and below (sample\#4 with 5.7~nm) the critical thickness. Whereas the monolayer graphene (sample\#5) was grown on a Si substrate, which at the same time functions as a homogeneous back gate.
 	}
 	\label{FigA1} 
 \end{figure}

HgCdTe/HgTe QWs samples~\#3 and \#4 were grown by molecular beam epitaxy on GaAs substrates with (013) orientation~\cite{Dvoretsky2010,Dvoretsky2017}.
%
%\re{Dvoretsky2016 = S.A. Dvoretsky, N.N. Mikhailov, I.V. Sabinina, G.Yu. Sidorov, Yu.G. Sidorov, V.A. Shvets, V.G. Remesnik, D.G. Ikusov, V.V. %Vasiliev, V.S. Varavin, M.V. Yakushev, A.V. Latyshev, and A.L. Aseev: MBE growth of HgCdTe het-ero- and nanostructures, Proc. of 18th Brazilian %Workshop on Semicond. Phys. (2016). }
%
The cross-section of these structures is shown in Fig.~\ref{FigA1}(b). To prevent strain-related effects due to lattice mismatch, a $5$~nm ZnTe layer and a 6~nm thick CdTe layer were grown on top of the substrate. The CdTe layer was followed by Hg$_{0.4}$Cd$_{0.6}$Te/HgTe/Hg$_{0.4}$Cd$_{0.6}$Te. The barrier thicknesses were 30~nm each. The QW thickness, $L_{\rm QW}$ was 8.1~nm in sample~\#3 and 5.7~nm in sample~\#4. In QWs of these thicknesses are characterized by parabolic dispersion with inverted and normal band order, respectively, see Refs.~\cite{Bernevig2006, Koenig2007, Olbrich2013}. The samples were square-shaped with the size 7$\times$7 mm$^2$ and were equipped with ohmic contacts at the corners. They were fixed in sample sockets and the contacts were connected to the socket pins via gold wires.

\begin{figure*}[t!]
	\centering \includegraphics[width=\linewidth]{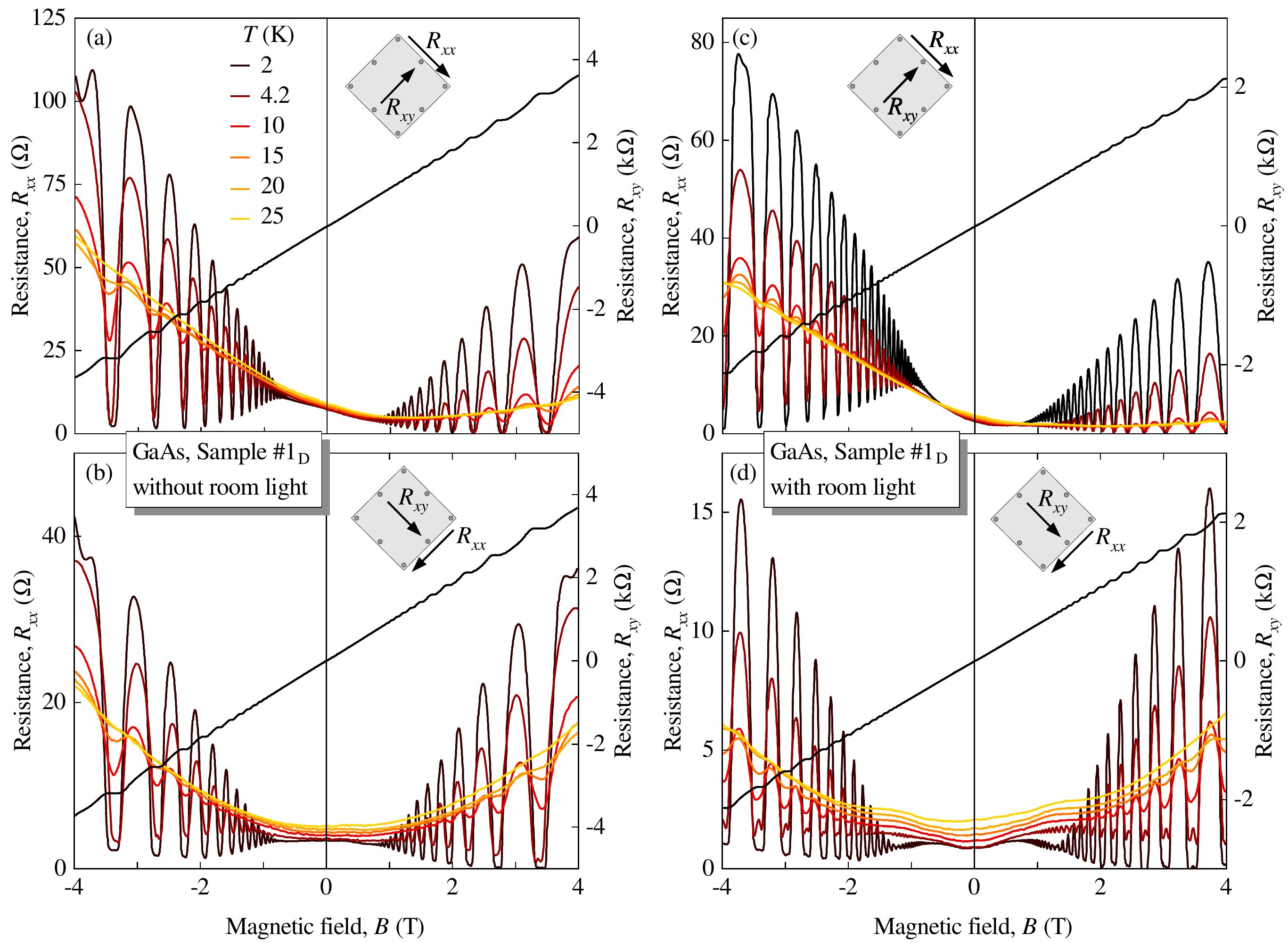}
	\caption{Hall resistance, $R_{xy}$, and temperature dependence of the longitudinal resistance, $R_{xx}$, measured for sample \#$1_{\rm D}$ for different vdP directions without [(a) and (b)] and with [(c) and (d)] room light illumination. The temperatures correspond to the traces following the indicated color code. The insets illustrate the measurement configuration.}
	\label{FigTr1}
\end{figure*}

\subsubsection{Encapsulated monolayer graphene}

%A hot pickup process similar to 1 was used to fabricate heterostructures. 

 %high quality hBN-encapsulated monolayer graphene (MLG) structure

The graphene sample~\#5 was a hBN-encapsulated monolayer graphene structure, for the cross-section see Fig.~\ref{FigA1}(c). First, the top hBN flake was picked up by a dome-shaped stamp made of polycarbonate film on polydimethylsiloxane. This hBN flake was then used to pick up single layer graphene and a bottom hBN flake was then picked up resulting in a hBN-graphene-hBN heterostructure. The stack was then placed on a standard p$^{++}$-doped Si/SiO$_2$ chip at a temperature of about \SI{180}{\degreeCelsius} for additional cleaning of the heterostructure interfaces. The Hall-bar structure was prepared by means of electron-beam lithography and reactive ion etching with O$_2$ and SF$_6$ gases. The hBN covering graphene at the contact areas of the Hall bar was subsequently etched away by SF$_6$ to enable surface contacts between graphene and 0.5~nm Cr / 100~nm Au circuit paths. To vary the carrier density, we used a back gate formed by the Si-substrate and 285 nm SiO$_2$ gate dielectric.

%\re{values for carrier density, mobility, scattime and thickness for GaAs 200 to 1000000 depending on dark/illum}
 
\newpage

\subsection{Transport and Magnetotransport}

\begin{figure*}[t!]
	\centering \includegraphics[width=\linewidth]{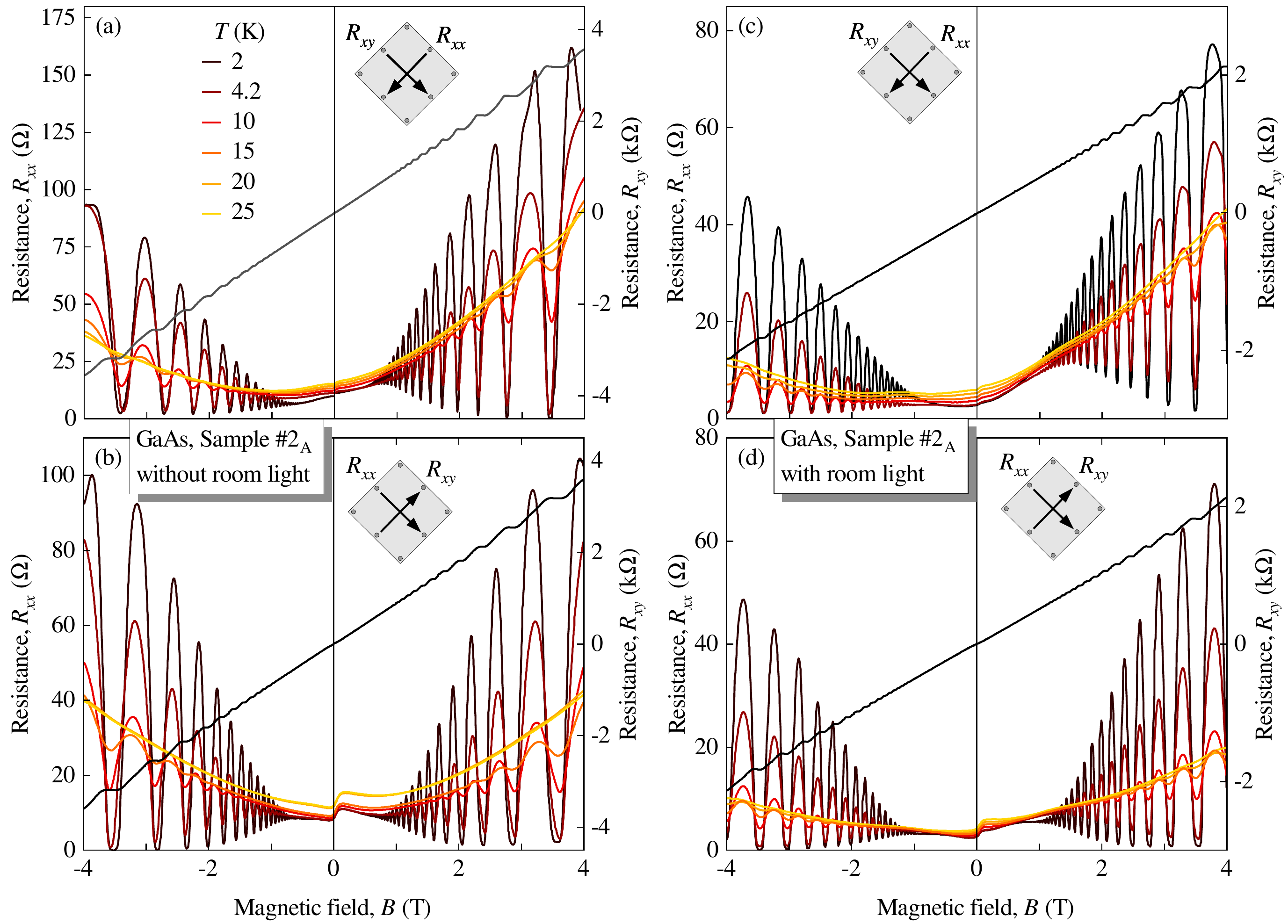}
	\caption{Hall resistance, $R_{xy}$, and temperature dependence of the longitudinal resistance, $R_{xx}$, obtained for sample\#$2_{\rm A}$ for different vdP directions without [(a) and (b)] and with [(c) and (d)] room light illumination. The temperatures correspond to the traces following the indicated color code. The insets illustrate the measurement configuration.}
	\label{FigTr3} 
\end{figure*}

\begin{figure}[t!]
	\centering \includegraphics[width=\linewidth]{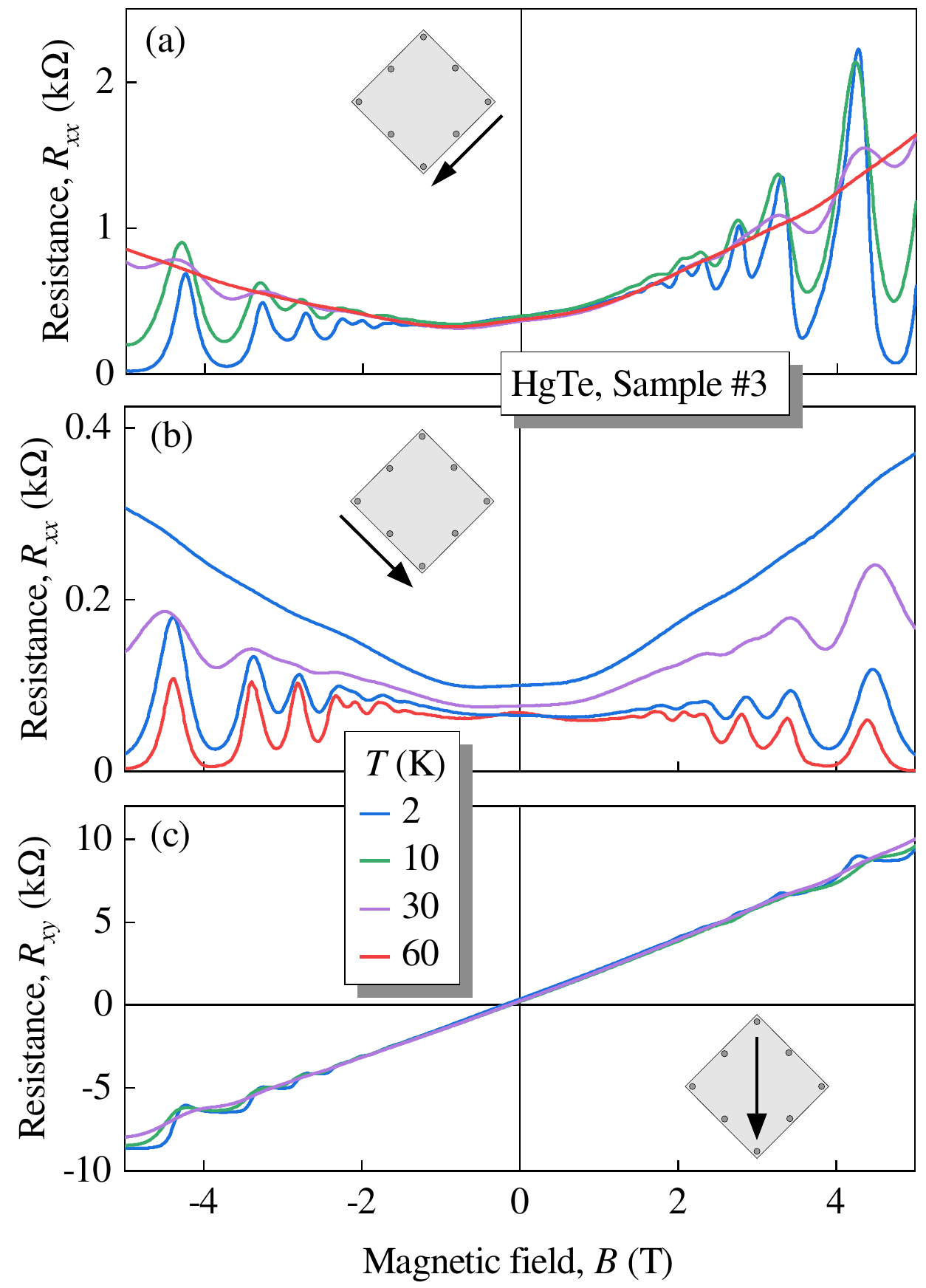}
	\caption{Temperature dependence of the magnetoresistance measured in 8.1~nm HgTe sample \#3. Panel (a) and (b) illustrate the longitudinal resistance, $R_{xx}$, obtained along two different directions of the vdP geometry. Panel (c) shows the Hall resistance, $R_{xy}$. The insets depict the corresponding measurement configuration.}
	\label{FigTr4} 
\end{figure}

\begin{figure}[t!]
	\centering \includegraphics[width=\linewidth]{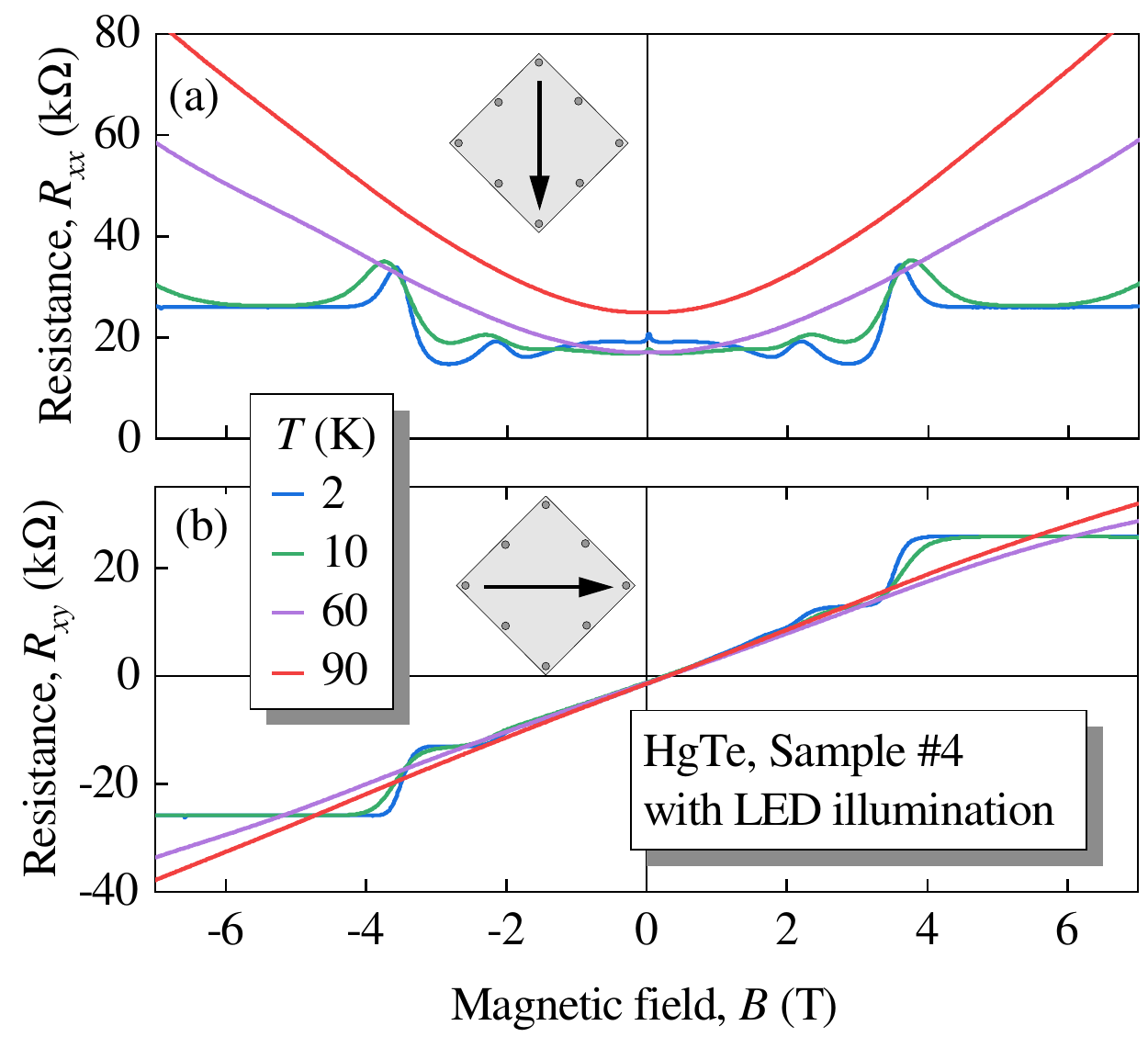}
	\caption{Temperature dependence of the magnetoresistance measured in 5.7~nm HgTe sample \#4 obtained after illuminating the sample with a red LED for approximately 15~mins. Panel (a) illustrates the two terminal longitudinal resistance, $R_{xx}$, while panel (b) shows the Hall resistance, $R_{xy}$. The insets depict the corresponding measurement configuration.}
	\label{FigTr5} 
\end{figure}

\begin{figure*}[t!]
		\centering \includegraphics[width=\linewidth]{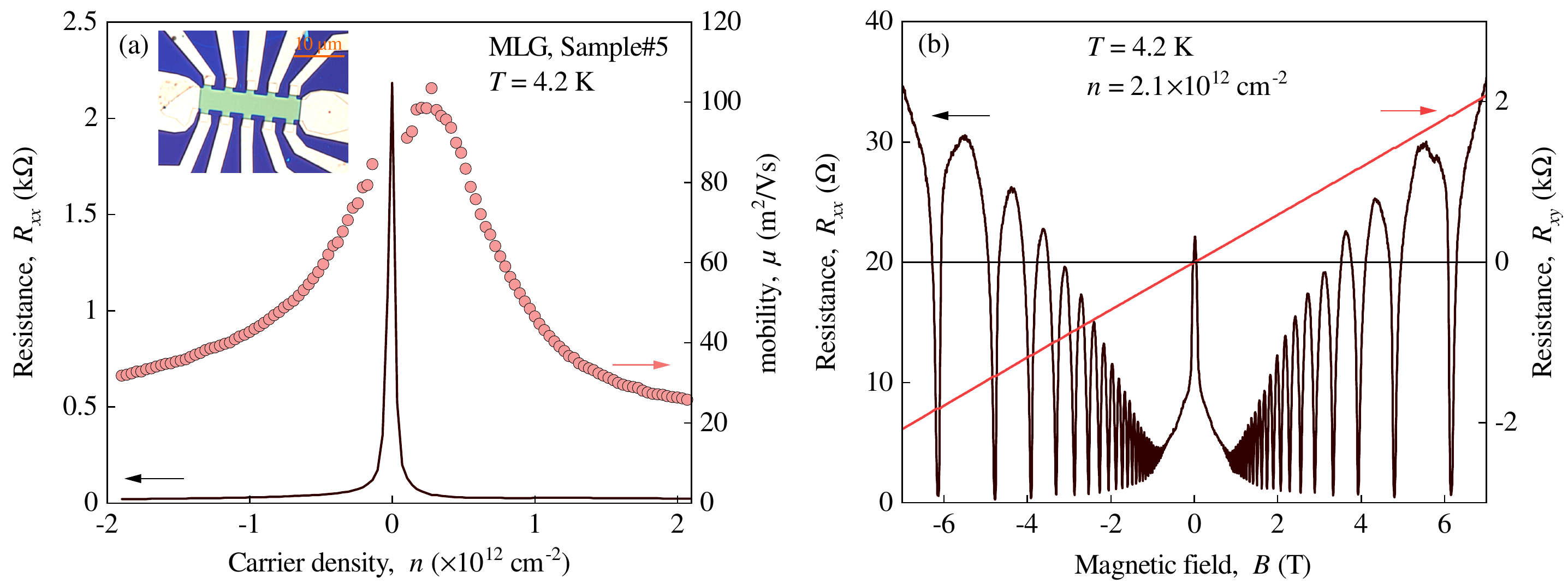}
	\caption{Panel (a): Longitudinal resistance (black curve) and mobility (red circles) as a function of the carrier density injected by a Si/SiO$_2$ back gate in monolayer graphene (sample \#5). A sharp peak in the resistance corresponds to the charge neutrality point. The inset illustrates a micrograph of the sample. Panel (b) depicts longitudinal (black) and Hall magnetoresistance (red) at a fixed carrier density of $n = 2.1\times 10^{12}$~cm$^{-2}$. $T = 4.2$~K. %\re{transconductance??? -> mobility}		
	}
	\label{FigTr6} 
\end{figure*}

The transport and magnetotransport results used to characterize the studied samples are presented for: (i) GaAs samples in Figs.~\ref{FigTr1} and \ref{FigTr3}, (ii) HgTe samples in Figs.~\ref{FigTr4} and \ref{FigTr5}, and (iii) graphene sample in Fig.~\ref{FigTr6}. The traces are given for the same temperatures as used in the photoresistance measurements. These data confirm the high quality of the studied materials. At low $T$ we observe strong SdHO in magnetotransport. Their modification at low $T$ by THz radiation due to heating of 2DES is the main effect in the photoresistance presented in main text and in %Sec.~\ref{add_results} 
the Supplemental Material.

%\re{Characterizations of the investigated sample were gathered by performing magnetotransport measurements. Figures~\ref{FigTr4} and \ref{FigTr5} show the results obtained for both samples. They were performed for three different temperatures of $T=2$, 10, 30 and 60~K. Panel a) and b) depict the longitudinal resistance $R_{\rm long}$ along the two different orientations of the edges of the sample. Shubnikov-de Haas oscillations (SdHO) can be observed in all cases, which are smeared at higher temperatures. This is a typical behaviour since the SdHOs are highly sensitive to temperature variations. The Hall resistance $R_{\rm Hall}$ measured across the sample is depicted in panel c). The data was used to determine the samples transport parameters, see \ref{Tab1}.}

\clearpage
\subsection{Methods}

To study the CR, the samples were placed in a temperature-regulated Oxford Cryomag optical cryostat with $z$-cut crystal quartz windows. For magneto-optical and magneto-transport experiments a magnetic field $B$ up to 7~T was applied normal to the 2DES plane. Most of the measurements were carried out, while keeping the samples in the dark. For that the cryostat windows were covered by black polyethylene films to avoid uncontrolled illumination by ambient light. In some cases the GaAs samples were illuminated by room light prior to measurements, which allowed to detect microwave-induced resistance oscillations (MIRO) coupled to harmonics of the CR, for review see~\cite{Dmitriev2012}. The corresponding magnetotransport data with and without illumination by room light is shown in Fig.~\ref{FigTr1}.

\begin{figure}[h!]
	\centering \includegraphics[width=\linewidth]{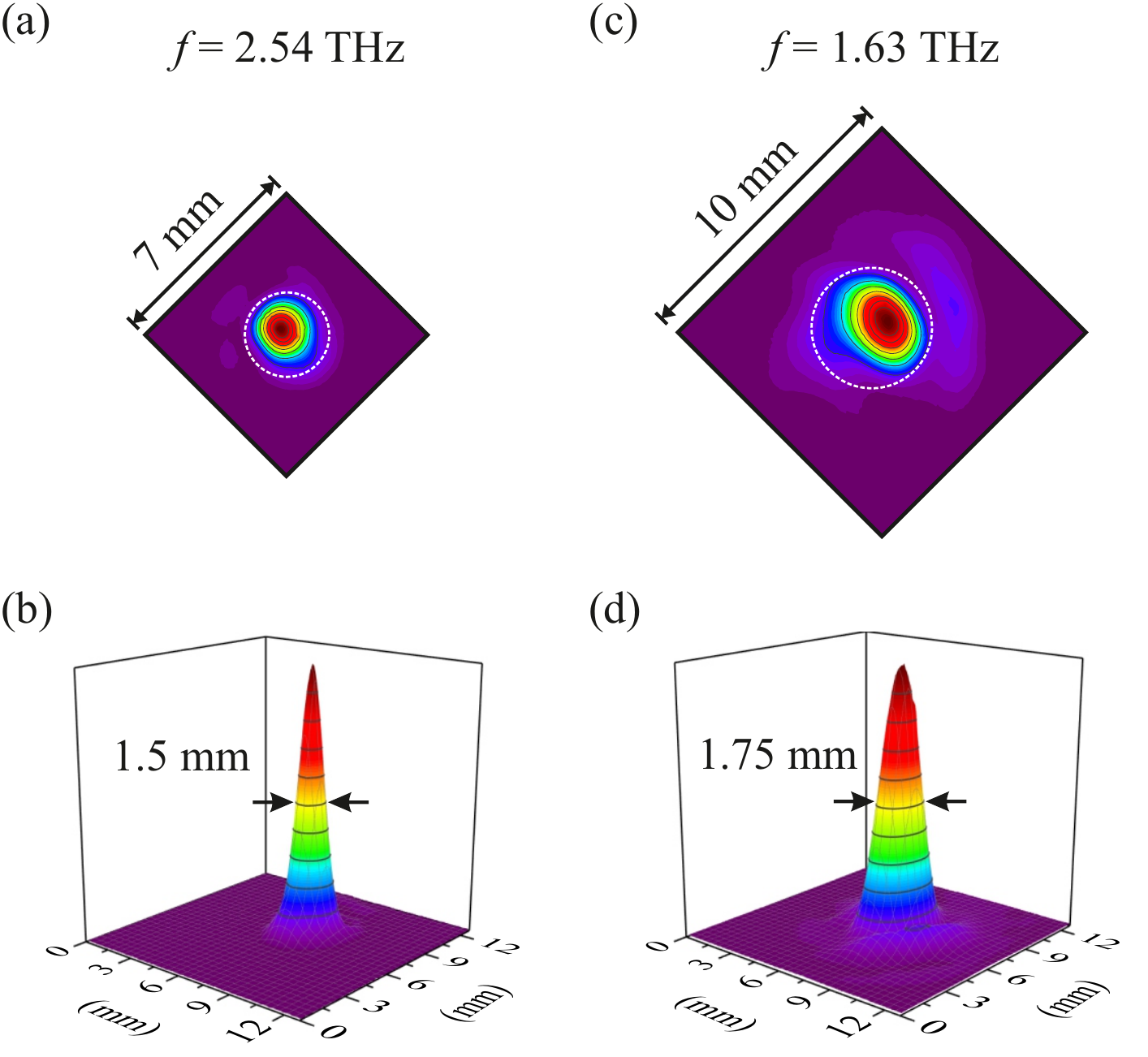}
	\caption{
		Intensity distribution of the focused laser spot at the sample's position for operation frequencies of $f = 2.54$ (a) and 1.63~THz (c), as captured with a pyroelectric camera. The laser mode is nearly Gaussian. The white dashed circles mark the diameter ($\approx$ 2.8~mm for $f = 2.54$ and $\approx$ 4.2~mm for 1.63~THz) for which the intensity drops to 1/e$^2$ ($\approx$ 13.5\%) of the peak intensity. Panel (b) and (d) show the corresponding 3D profile of the focused laser spot at the sample's position. The full width at half maximum spot diameter, labeled by horizontal arrows, is approximately 1.5~mm (1.75~mm) for $f = 2.54$~THz (1.63~THz).}
	\label{FigA2} 
\end{figure}

Photoresistance measurements were performed across the sample in two-terminal configuration as depicted in Fig.~\ref{FigS1}, using ohmic contacts at the corners. The response to the modulated THz radiation was measured using a standard lock-in technique. The modulation at $f_{\rm chop} = 140$~Hz was obtained by an optical chopper. Two types of measurements applying either $dc$ or $ac$ bias were used. In the former a $dc$ bias, $U_{\rm dc}$, was applied over a load resistor, $R_{\rm L}$, to the sample. The photoresistance was extracted by subtracting the two measured signals obtained for opposite polarities of the $dc$ bias.  Using that, by definition, the photoconductive signal has an opposite sign for positive and negative bias, we extracted the photoconductive contribution $V_{\rm pc}$ as an odd part of $V(U_{\rm dc})$,
\begin{equation}
\label{eqodd}
V_{\rm pc} = \frac{V(U_{\rm dc})-V(-U_{\rm dc})}{2}.
\end{equation}
This procedure allowed us to extract photoresistance from the total signal including possible photocurrent contributions. 
The second configuration uses an alternating bias voltage of low frequency, $f_{\rm ac}$ = 7.757~Hz, over a load resistor to the sample, while the THz-radiation is simultaneously modulated by an optical chopper with $f_{\rm chop} = 140$~Hz. This results in a so-called double-modulation and requires a series connection of two lock-in amplifiers for signal extraction.%Such a so-called double-modulation technique is possible by connecting two lock-in amplifiers in series. 
The first one is phase-locked to the chopper frequency filtering the total signal. The output consists of two distinct photosignals: a constant component proportional to the photocurrent and a component modulated at $f_{\rm ac}$, which is proportional to the photoconductivity. This output feeds the input of the second lock-in amplifier phase-locked to $f_{\rm ac}$ providing a voltage determined by the magnitude of the photoconductivity signal $V_{\rm pc}$ (see also, e.g., Ref.~\cite{Kozlov2011, Otteneder2018}).
%\re{Is already introduced above: Otteneder2018 = 2018 98 245304 Sign-alternating photoconductivity Otteneder 292.pdf}
%\item  M.\,Otteneder, I.\,A.\,Dmitriev, S.\,Candussio, M.\,L.\,Savchenko, D.\,A.\,Kozlov, V.\,V.\,Bel'kov, Z.\,D.\,Kvon, N.\,N.\,Mikhailov, S.\,A.\,Dvoretsky, and S.\,D.\,Ganichev,\\ \textit{Sign-alternating photoconductivity and magnetoresistance oscillations induced by terahertz radiation in HgTe quantum wells,} Phys. Rev. B \textbf{98}, 245304 (2018).
 The photoresistance $\Delta R$ in both measurement configurations is related to the measured signal $V_{\rm pc}$ as $\Delta R = V_{\rm pc}R_{\rm L}/\abs{U_{\rm dc}}$. Simultaneously to the photoresistance measurements, the transmitted radiation was recorded by a low-noise pyroelectric detector.  The large-sized vdP samples, except graphene, excluded any influence of effects originating from the illumination of edges and contacts. The sample size was chosen to be significantly larger than the THz spot size. In Fig.~\ref{FigS1}(b) and Fig.~\ref{FigA2}(a) and (c) we compare the sample's dimensions (square shaped with a side length of 10~mm for GaAs and 7~mm for HgTe) and the spot sizes for the corresponding radiation frequencies.

\clearpage

%voltages $V_x^\text{dc}$ whereas the the photogalvanic current should be insensitive to the polarity of $V_x^\text{dc}$,

 %phase-locked to the chopper modulation frequency of the THz-radiation

 %for both $dc$-current polarities in sequence. Using 

%, using either a $dc$- or a $ac$-current bias. In the former case a $dc$-current, $I_{dc}$, was applied over a load resistor, $R_{\rm L}$, to the sample.  as a voltage drop, $V$, using a standard lock-in technique phase-locked to the chopper modulation frequency of the THz-radiation

%Using that, by definition, the linear-in-$V_x^\text{dc}$ photoconductivity signal should have opposite sign for positive and negative bias voltages $V_x^\text{dc}$ whereas the the photogalvanic current should be insensitive to the polarity of $V_x^\text{dc}$, we extracted the photoconductivity contribution $U_{\rm pc}^{\rm circ}$ as an odd part of $U_y^{\rm circ}(V_x^\text{dc})$,
%
%\begin{equation}
%\label{eqodd}
%U_{\rm pc}^{\rm circ}= \frac{U_y^{\rm circ}(V_x^\text{dc})-U_y^{\rm circ}(-V_x^\text{dc}) }{2}.
%\end{equation}
%The photoresistance was extracted by subtracting the two measured signals obtained for opposite polarities of the $dc$ bias. 

%A \textit{dc} bias voltage $V_x^\text{dc}$ is applied in $x$ direction along the Hall bar. The transversal photosignal $U_y$ is picked up as a voltage drop over load resistors $R_{\rm L} = 50 \,\Omega$. Subtracting the photosignal for opposite $V_x^\text{dc}$ polarities yields the photoconductivity signal $U_{pc}=[U_y(V_x^\text{dc})-U_y(-V_x^\text{dc})]/2$.

\begin{figure*}[h!]
	\centering \includegraphics[width=\linewidth]{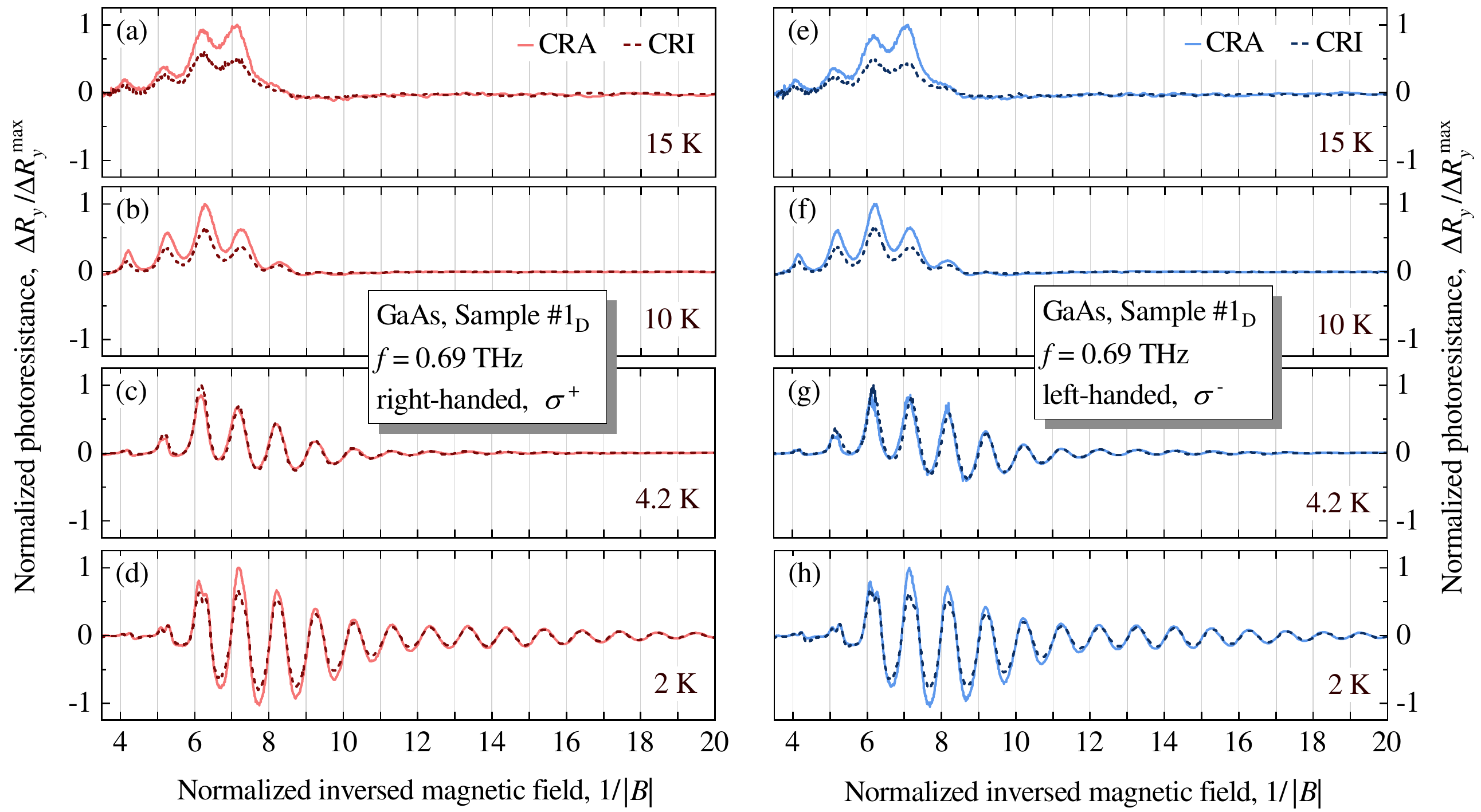}
	\caption{Photoresistance vs. absolute value of the inversed magnetic field for GaAs sample~\#1$_{\rm D}$ obtained for right- [(a) to (d)] and left-handed [(e) to (h)] radiation helicities at different temperatures. The photoresistance $\Delta R_y$ is normalized to its maximum value, $\Delta R^{\rm max}_y$ on the CR-active (CRA) side, except the traces at $T = 4.2$~K, where the maximum value lies at the CR-inactive (CRI) polarity. For a clear presentation the inversed magnetic field scale was multiplied for each temperature by the difference of one period of the photoresistance oscillations at $T = 2$~K. The traces for the CRA (solid lines) and CRI (dashed lines) magnetic field polarities are plotted together for each temperature.}
	\label{FigA6} 
\end{figure*}

\subsection{Analysis of SdHO-related photoresistance oscillations}

\begin{figure*}[t!]
	\centering \includegraphics[width=\linewidth]{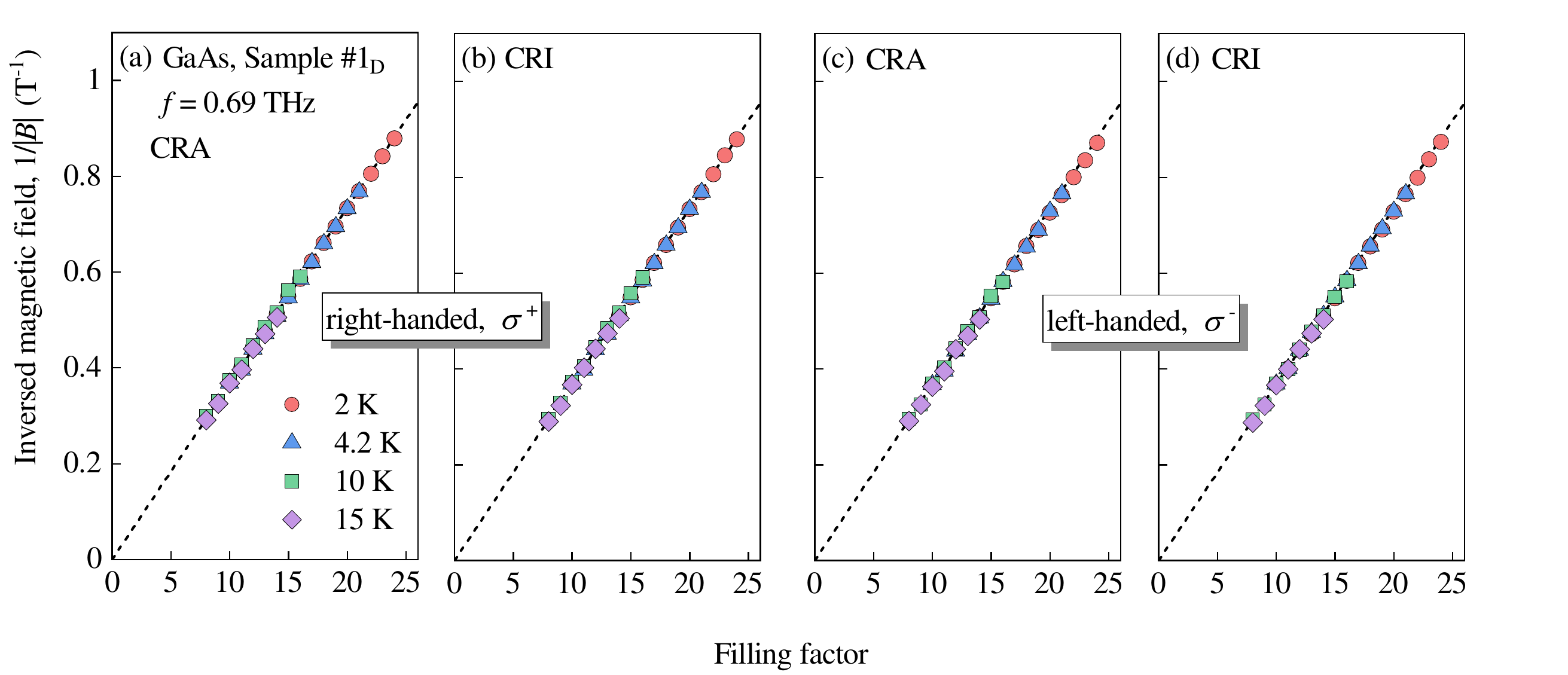}
	\caption{The inversed magnetic field positions of the oscillation's extrema plotted against the filling factor of the Landau levels.}
	\label{FigA7} 
\end{figure*}

To analyse the oscillations in the photoresistance traces we replotted the results exemplarily for the GaAs sample~\#1$_{\rm D}$ against the inversed magnetic field. Figure~\ref{FigA6} shows the replotted data for $f = 0.69$~THz and temperatures up to $T = 15$~K. Here, panels (a) to (d) correspond to right-handed, whereas panel (e) to (h) to the left-handed helicity. For clarity the inverted magnetic field is rescaled to the period of photoresistance oscillations at $T = 2$~K. This presentation displays that the photoresistance oscillations are indeed 1/$B$-periodic and behave similarly in the whole temperature range, which is clearly demonstrated by the equidistant gray guide lines. This is additionally supported by fact that the inverted magnetic field positions of the oscillation's extrema plotted against the filling factor of the Landau levels follow a straight line crossing the origin, see Fig.~\ref{FigA7}. Note that, as opposed to the SdHO in the dark magnetotransport, the photoresistance oscillations exhibit maxima (minima) at even (odd) Landau level filling factors. 

%In addition, the period of the SdHO in the magnetotransport measurements is linked to the carrier density, which is routinely determined from the Hall measurements. Applying the latter to the oscillations, which emerges in the photoresistance, we found that their period reflects the carrier density of the sample confirming, all together, that these oscillations are unambiguously related to the SdHO in conventional magnetotransport measurements. 

%\subsection{MIRO analysis}

\clearpage

\subsection{Further examples of experimental results obtained at higher frequency and different samples}
\label{add_results}

\begin{figure*}[h!]
	\centering \includegraphics[width=\linewidth]{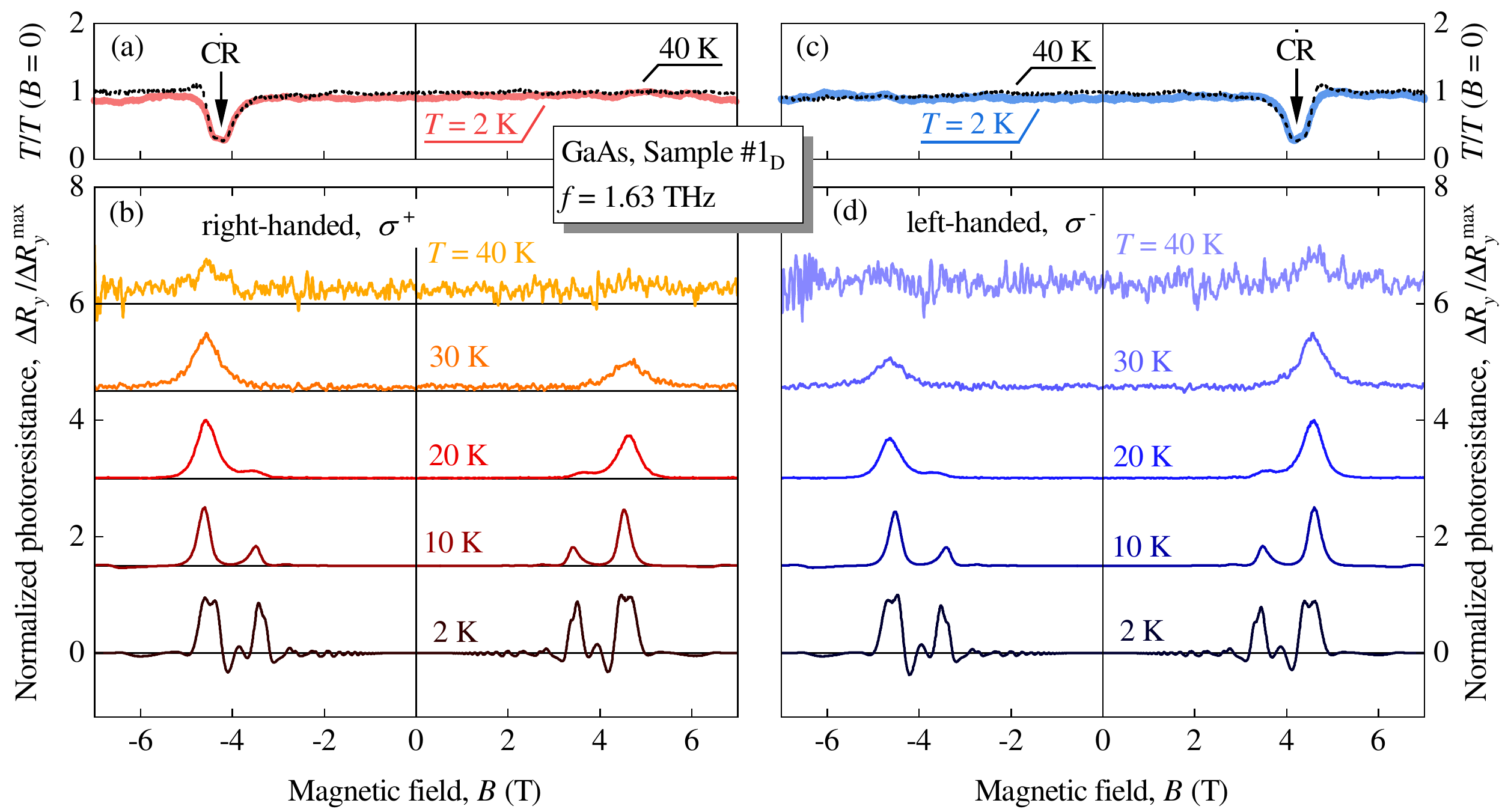}
	\caption{Transmitted signals measured on the GaAs sample\#1$_{\rm D}$ for both right- and left-handed radiation helicities, illustrated in panels (a) and (c), respectively. The traces were recorded for $T = 2$ and 40~K as a response to radiation at frequency $f = 1.63$~THz and are normalized to the value at $B = 0$. Vertical arrows labeled CR show the position of the cyclotron resonance. Panels (b) and (d) show the corresponding photoresistance, $\Delta R_y$, normalized to its maximum value, $\Delta R^{\rm max}_y$. The traces were recorded for a set of temperatures ranging from $T = 40$~K down to $T = 2$~K. Each trace is shifted for clarity by $\Delta R_y$/$\Delta R^{\rm max}_y$ = 1.5.}
	\label{FigA8} 
\end{figure*}

\begin{figure*}[h!]
	\centering \includegraphics[width=\linewidth]{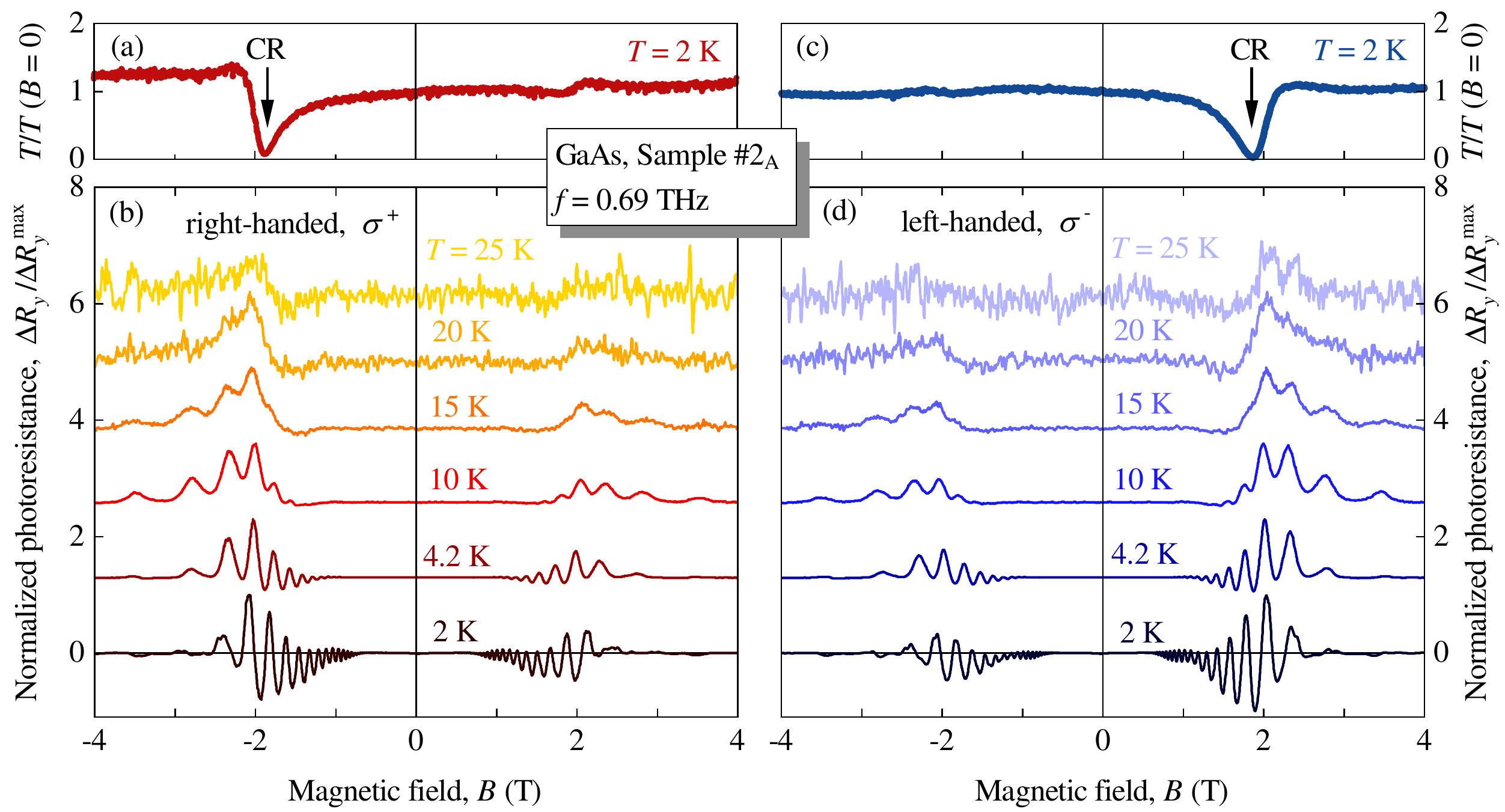}
	\caption{Transmitted signals measured on the GaAs sample\#2$_{\rm A}$ for both radiation helicities, illustrated in panels (a) and (c), respectively. The traces were recorded at $T = 2$~K for $f = 0.69$~THz radiation and are normalized to the value at $B = 0$.  The vertical arrows indicate position of CR. The corresponding photoresistance traces, $\Delta R_y$, are shown in panels (b) and (d) and are normalized to their maximum value, $\Delta R^{\rm max}_y$. The curves were measured for a set of temperatures ranging from $T = 25$~K down to $T = 2$~K. Each trace is shifted for clarity by $\Delta R_y$/$\Delta R^{\rm max}_y$ = 1.5.}
	\label{FigA9} 
\end{figure*}

\begin{figure*}[h!]
	\centering \includegraphics[width=\linewidth]{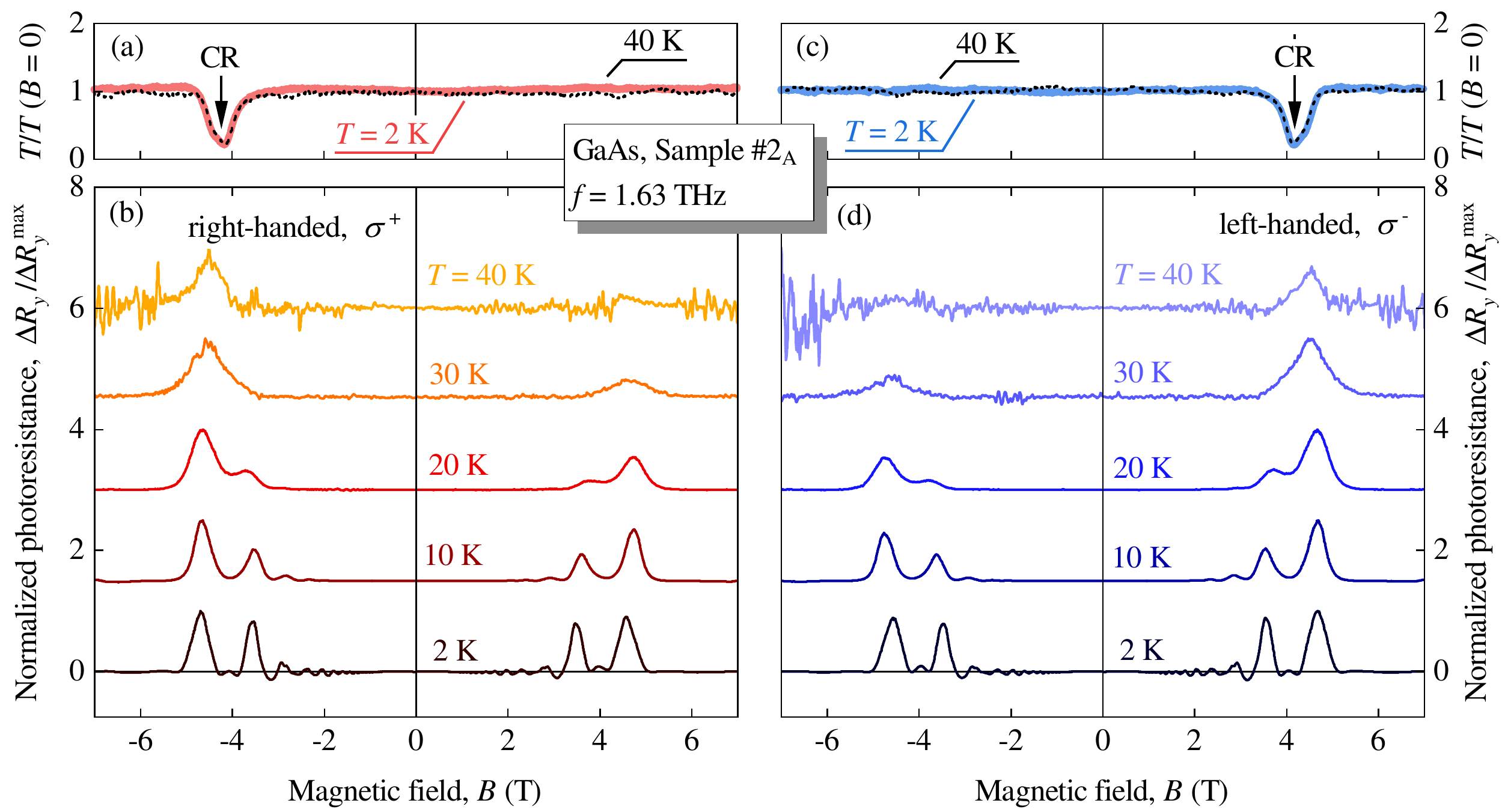}
	\caption{Panels (a) and (c): Transmitted signals measured on the GaAs sample \#2$_{\rm A}$ for right- and left-handed radiation. The traces were recorded at $T = 2$ and 40~K under $f = 1.63$~THz radiation and were normalized to the value at $B = 0$.  The vertical arrows labeled with CR mark the cyclotron resonance. Panels (b) and (d): The corresponding photoresistance, $\Delta R_y$, normalized to its maximum value, $\Delta R^{\rm max}_y$, shown for a set of temperatures ranging from $T = 25$~K down to $T = 2$~K. Each trace is shifted for clarity by $\Delta R_y$/$\Delta R^{\rm max}_y$ = 1.5.}
	\label{FigA10} 
\end{figure*}

\begin{figure*}[h!]
	\centering \includegraphics[width=\linewidth]{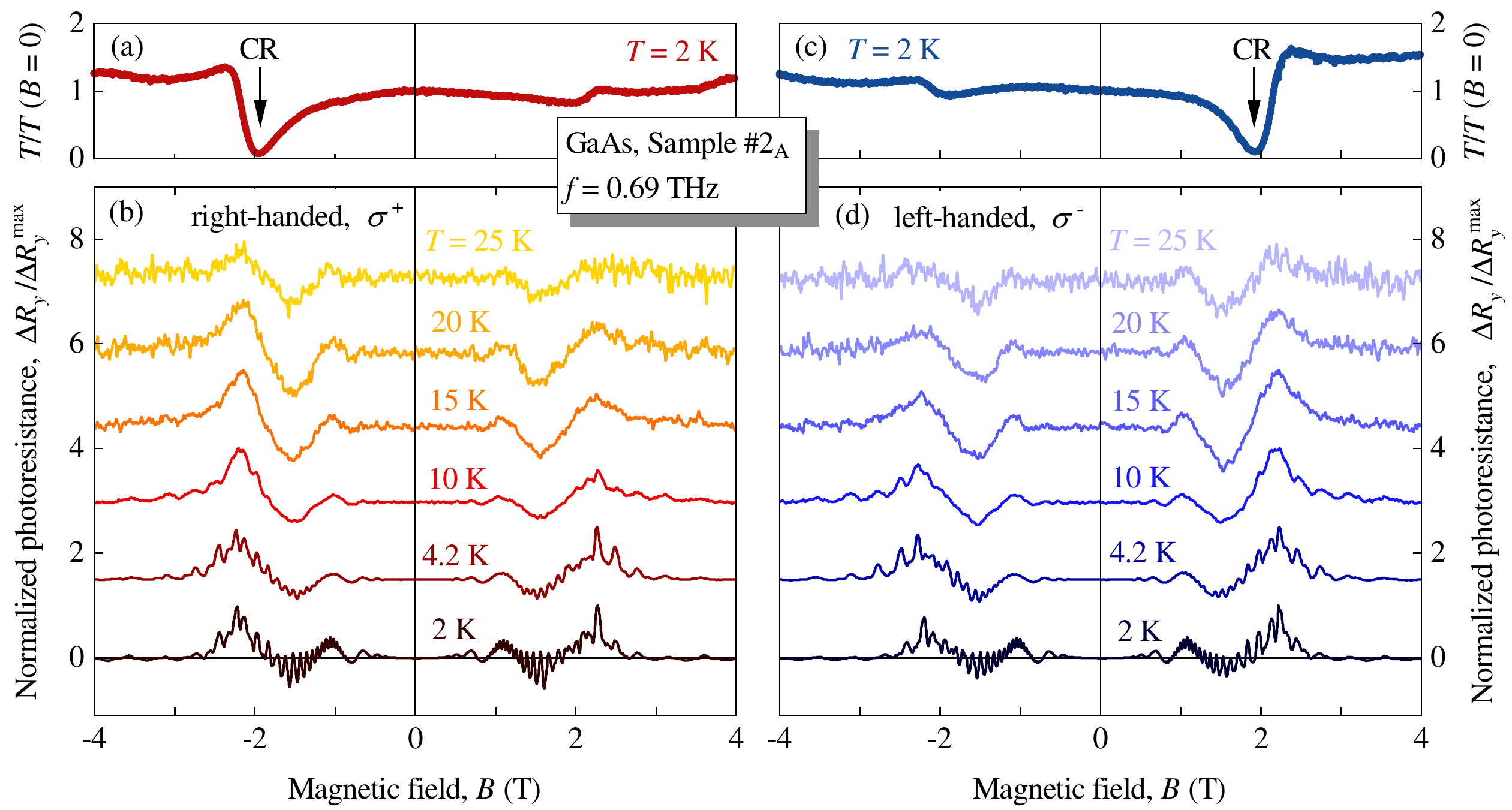}
	\caption{Panels (a) and (c): Transmitted signals measured on the GaAs sample \#2$_{\rm A}$ for right- and left-handed radiation after exposing the sample to room light illumination. The traces were recorded at $T = 2$~K for $f = 0.69$~THz and were normalized to the value at $B = 0$.  The black vertical arrows labeled with CR depict the cyclotron resonance. Panels (b) and (d): Corresponding photoresistance, $\Delta R_y$, normalized to its maximum value, $\Delta R^{\rm max}_y$, shown for a set of temperatures ranging from $T = 25$~K down to $T = 2$~K. Each trace is shifted for clarity by $\Delta R_y$/$\Delta R^{\rm max}_y$ = 1.5.}
	\label{FigA11} 
\end{figure*}

\begin{figure}[h!]
	\centering \includegraphics[width=\linewidth]{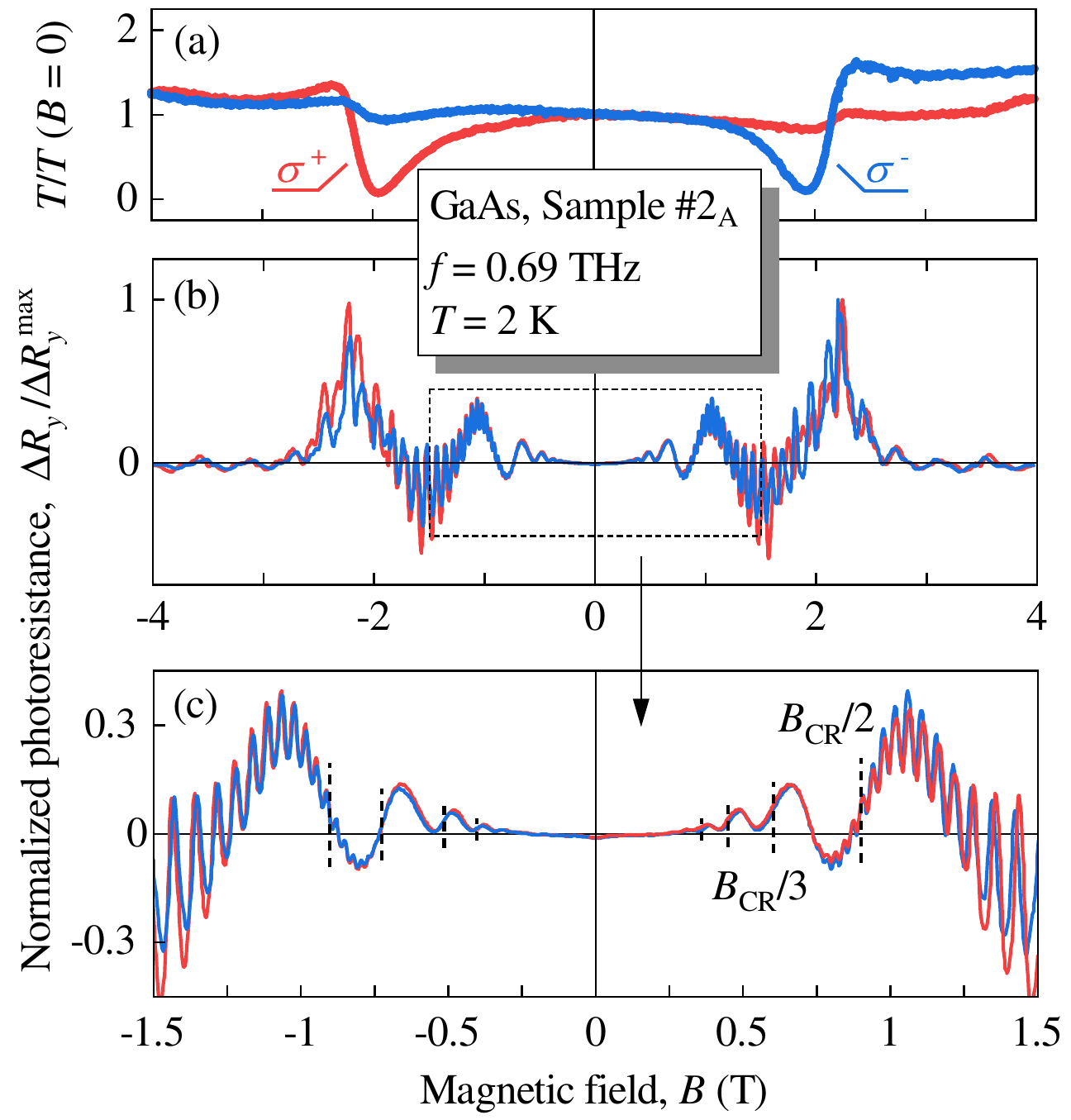}
	\caption{Transmitted signal (a) and the corresponding photoresistance (b) obtained at $T = 2$~K under $f = 0.69$~THz radiation on GaAs sample\#2$_{\rm A}$ after exposure to room light. The red (blue) traces were measured under right- (left-) handed radiation. Panel (c) shows a zoom-in of the low-$B$ area marked by a dashed rectangular in panel (b). Dashed vertical lines mark the nodes of MIRO.}
	\label{FigA12} 
\end{figure}

\begin{figure}
	\centering \includegraphics[width=\linewidth]{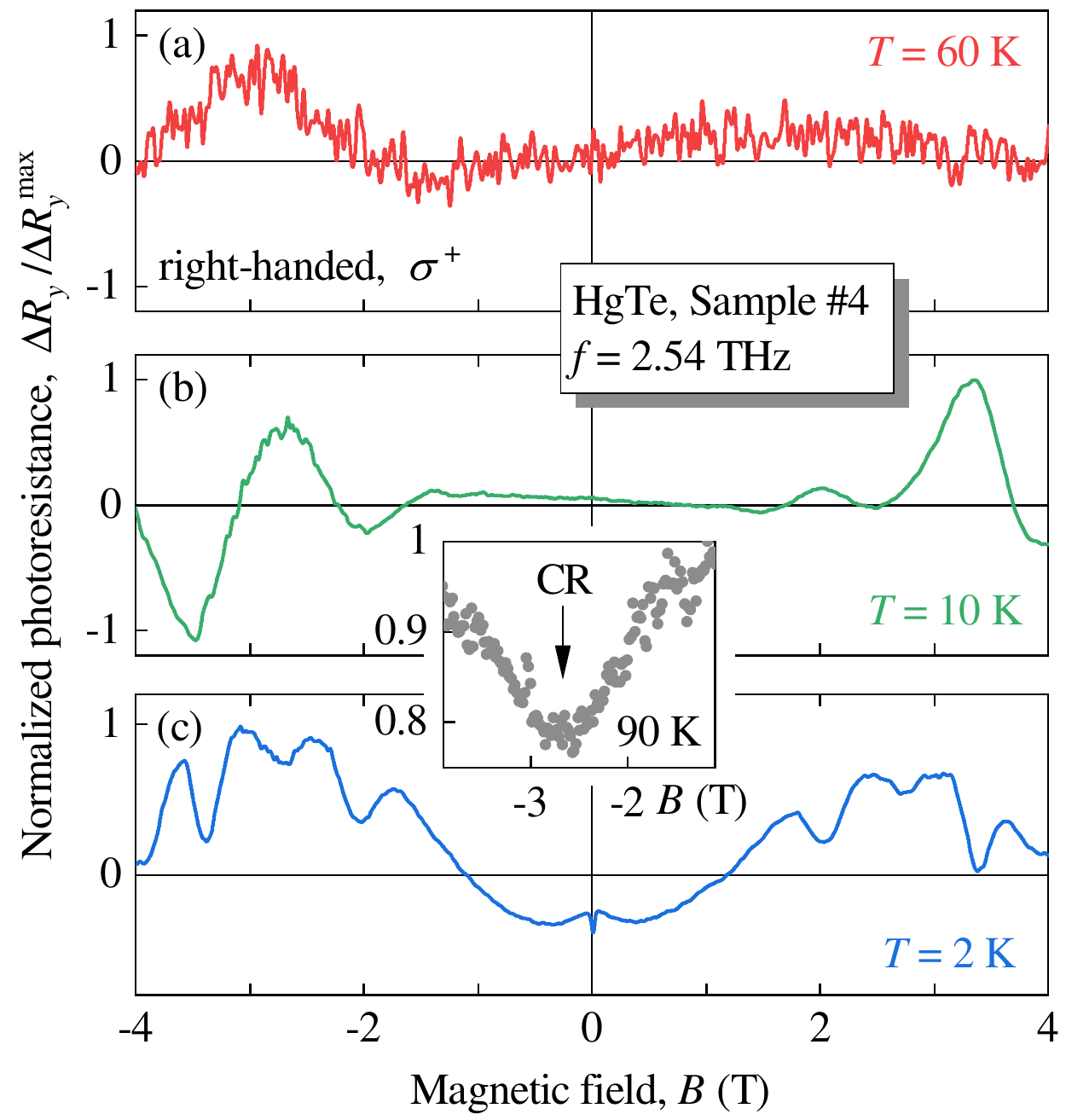}
	\caption{Normalized photoresistance $\Delta R_y$/$\Delta R^{\rm max}_y$ measured under $f = 2.54$~THz right-handed radiation on HgTe sample \#4 at $T = 60$, 10 and 2~K, panels (a), (b) and (c), respectively. The inset shows CR in the corresponding transmission data obtained at $T = 90$~K. }
	\label{FigA13} 
\end{figure}

In addition to the results discussed in the main text, Fig.~\ref{FigA8} illustrates a temperature set for the GaAs sample~\#1$_{\rm D}$ measured at a higher radiation frequency of $f = 1.63$~THz. Similar traces were also obtained for GaAs sample~\#2$_{\rm A}$ at $f = 0.69$ and 1.63~THz, see Figs.~\ref{FigA9} to \ref{FigA12}, and for HgTe sample~\#4 at $f = 2.54$~THz, see Fig.~\ref{FigA13}. All results presented here are in a good agreement with those selected for presentation in the main text and support the conclusions drawn there.

\clearpage

%\end{document}

%\bibliography{references.bib}
%\bibliography{all_lib}
%\bibliography{all_lib.bib}
%\IfFileExists{I:/_Papers_Abstracts/JabRef_Bibliography/all_lib.bib}{\bibliography{I:/_Papers_Abstracts/JabRef_Bibliography/all_lib.bib}}{\bibliography{all_lib}}

\end{document}